# Turbulent wind thrust control to reduce pitch motion of floating wind platforms with improved rotor operation

Yi Zhang, Peter Stansby*, David Apsley, Hannah Mullings

School of Engineering, The University of Manchester, Manchester M13 9PL, UK

## Abstract

For the dynamic analysis of floating offshore wind turbines (FOWTs) in realistic operating environment, this paper develops a coupled aero-hydro-mooring-servo model applicable to turbulent wind and irregular sea states with high computational efficiency. A modified rotor control strategy with platform motion feedback is proposed with a novel gain-scheduling technique to mitigate the negative damping effect on platform motions and decouple the rotor dynamics from the platform dynamics for better rotor operating performance. Firstly, the performance for the bottom-fixed wind turbine in steady uniform wind is validated and the turbulent wind tests show that using the rotor-disk-averaged wind speed for control is beneficial for reducing fluctuations of wind thrust and operational parameters compared with using the hub-height wind speed. For the FOWT, the negative damping phenomenon at above-rated wind speeds when using the baseline control is demonstrated through a range of wind and wave scenarios, and effects of control strategy and turbulent wind are investigated. The results indicate that the modified control strategy eliminates the negative damping effect on the platform pitch while maintaining small variations in rotor speed. Though the control including the blade pitch compensation from platform motion feedback with a constant gain can also eliminate the negative damping effect, it produces much larger fluctuations in rotor speed and causes larger overspeed exceeding the safety threshold of 20% at large wind speeds. Compared to steady wind, turbulent wind yields significantly larger low-frequency platform responses and increases the maximum rotor speed by 7.9% to 23.7% for the wind speeds considered.

## 1. Introduction

Offshore wind energy is gradually becoming one of the major sources of electricity supply worldwide to achieve the goal of net zero emissions by 2050. It is predicted that the total installed capacity of global offshore wind turbines will increase to 228 GW by 2030 and reach nearly 1000 GW by 2050 (Irena, 2019). As offshore wind energy rapidly advances, the rated power capacity of individual wind turbines has risen from 5 MW to 15 MW and the development has gradually expanded from nearshore to deep sea areas over the past two decades. A variety of floating offshore wind turbine (FOWT) concepts have been proposed and put into practical use (Edwards et al., 2023, 2024; Zeng et al., 2024). Floating wind platforms can generally be categorized into four main types: semi-submersible, spar, tension leg and barge, but some novel concepts may fall outside these categories. There are several representative FOWTs projects in recent years, such as Hywind Scotland, Fukushima FORWARD, Windfloat Atlantic, VolturnUS, TetraSpar, Hywind Tampen and so on (Edwards et al., 2023; Zhang et al., 2024). As extensively discussed in van Kuik et al. (2016) and Veers et al. (2023), there are grand challenges in the design, manufacture and operation for large-scale modern wind turbine systems, particularly for FOWTs operating in complex atmospheric and marine environments. The reduction in levelized cost of energy (LCOE) of FOWTs is the main design goal in offshore wind energy, and efforts are still made through various means to improve the survivability, reliability and safe accessibility of FOWTs.

Modelling the dynamic characteristics of FOWTs is a complex task, which requires a comprehensive consideration of factors such as platform hydrodynamics, turbine blade aerodynamics, structural dynamics, mooring system and control strategy. Several popular mid-fidelity FOWT analysis tools now available include Bladed, HAWC2, OpenFAST, OrcaFlex (Otter et al., 2022; Subbulakshmi et al., 2022). Validation within the OC3, OC4, and OC5 projects has confirmed the reliability of these modeling tools (Jonkman et al., 2010; Robertson et al., 2014, 2017) through code-to-code comparisons. The classical blade element momentum theory (BEMT) is generally used for aerodynamic modelling in these mid-fidelity models due to its high efficiency. Recent studies employ high-fidelity computational fluid dynamics (CFD) models for FOWT dynamic analysis, but their computational costs are invariably very high. For these CFD models, the aerodynamic modelling can be implemented by using a blade-resolved model (Tran and Kim, 2016; Liu et al., 2017) requiring significant computational cost and complex dynamic mesh technique, or the actuator disk method (ADM) (Mikkelsen, 2004; Réthoré et al., 2014) and actuator line method (ALM) (Sørensen and Shen, 2002; Apsley and Stansby, 2020; Arabgolarcheh et al., 2022) as a computationally efficient alternative. The CFD applications in aerodynamics, hydrodynamics and fully coupled aero-hydrodynamics of FOWTs have been reviewed in Zhang et al. (2024). Although FOWT dynamics have been extensively investigated numerically and experimentally (see Subbulakshmi et al., 2022), there are still a range of problems to be addressed, such as the robust control and co-design for FOWTs, atmospheric turbulence effect and wake flow characteristics.

To mitigate the impact of environmental loads and ensure the safe operation of the wind turbine, control systems are usually employed. The control system has a crucial impact on the dynamic response and power output of FOWTs. The baseline controller for land-based wind turbines is usually based on the conventional proportional–integral (PI) control on the generator torque and blade pitch (Njiri and Söffker, 2016; Abbas et al., 2022). The basic DTU Wind Energy controller (Hansen and Henriksen, 2013) and the Reference Open-Source Controller (ROSCO) (NREL, 2021) are the two reference controllers representing industry standards. The rotor control for FOWTs is more complex and challenging due to the dynamic coupling between the floating platform and the wind turbine. One of the notable issues is the platform/tower motion instability if directly applying the baseline controller designed for land-based wind turbines to FOWTs at above-rated wind speeds, which is often referred to as "negative damping" (Nielsen et al., 2006; Larsen et al., 2007; Jonkman, 2008) in this field. This is essentially a control-induced instability of floating wind platforms and can be briefly explained as follows. During above-rated operation when the FOWT moves backward in the direction of wind, the relative wind speed coming into the rotor plane decreases and the baseline blade pitch controller will reduce the blade pitch angle to maintain rated rotor speed. This causes an increase in thrust and drives the FOWT backward further. Similarly, the thrust decreases when the FOWT moves forward. This can lead to negatively damped oscillations of the FOWT system, with the platform instability and poor rotor performance. This negative damping phenomenon has attracted considerable attention. A simple method to mitigate the negative damping effect is to reduce the controller gains while retaining the baseline controller architecture, commonly known as "detuning" (Larsen et al., 2007; Fleming et al., 2014). In this method, the steady-state operating point is still determined based on the inflow wind speed without considering the relative speed due to platform motion. Though the detuning strategy is generally beneficial, it may inadequately stabilize the platform at near-rated wind speeds. A large detuning also leads to a degradation in generator speed and power regulation, and increases the risk of generator overspeeding (Yu et al., 2018; Stockhouse et al., 2022). Another approach is adding floating feedback loops (e.g., van der Veen et al., 2012; Fischer, 2013; Yu et al., 2018; Stockhouse et al., 2022) to compensate for the coupling between the rotor dynamics and platform dynamics, which forms a multi-input, multi-output (MIMO) controller. Several advanced controllers are also designed for FOWTs, such as the linear quadratic regulator (LQR) (Namik and Stol, 2010, 2011; Lemmer et al., 2016), disturbance accommodating control (DAC) (Namik and Stol, 2011; Menezes et al., 2018), model predictive control (MPC) (Raach et al., 2014; Lemmer et al., 2015) and feedforward control incorporating wind and wave forecasting techniques (Schlipf et al., 2015; Scholbrock et al., 2016; Fontanella et al., 2021; Russell et al., 2024). A range of control technologies for FOWTs have been reviewed in Namik and Stol (2013), Shah et al. (2021), López-Queija et al. (2022) and Stockhouse et al. (2023). Most research on FOWTs with rotor control is implemented based on potential flow-based models. To our knowledge, there are only a few CFD studies (Quallen and Xin, 2016; Yang et al., 2023; Huang et al., 2025) and experimental studies (Goupee et al., 2014, 2017; Madsen et al., 2020; Meng et al., 2023) incorporating rotor control for FOWT dynamic analysis.

In this study, a coupled aero-hydro-mooring-servo model is developed for dynamic modelling of the IEA Wind 15 MW reference turbine with semi-submersible platform. We integrate the new aerodynamic model OREGEN_BEMT, hydrodynamic model OREGEN_X, mooring model, and multiple options of rotor control algorithms. The mooring model can be the quasi-static mooring or the dynamic lumped-mass model MoorDyn (Hall and Goupee, 2015). In this paper, we use a more general variant of the baseline controller that sets the operating point based on instantaneous relative wind speed. However this may also lead to platform instability and power loss due to the dynamic coupling between the platform and turbine, through negative damping phenomenon. This is improved by introducing feedback from platform motion to both blade pitch and generator torque actuators. A modified control strategy is proposed with a novel gain-scheduling technique for platform motion feedback to mitigate the negative damping effect on platform motions while improving the rotor operating performance. The proposed coupled model is employed in dynamic simulations of the FOWT with different control strategies across a range of wind and wave scenarios, including realistic turbulent wind and irregular sea states. By analysing the platform motions, wind thrust, blade pitch angle and rotor speed, the negative damping phenomenon is demonstrated and the mitigating effects of control strategy on the platform response and rotor performance are examined under extensive operational conditions.

This paper is structured as follows. Section 2 formulates the proposed coupled aero-hydro-mooring-servo model for FOWT simulations. Section 3 describes the FOWT model and lists the key properties and numerical set-ups. In Section 4, we investigate the effects of turbulent wind and different control strategies on the platform response and rotor performance under diverse wind and wave scenarios. The conclusions of this study are summarized Section 5.

## 2. Mathematical formulation

In this study, we perform a coupled aero-hydro-mooring-servo simulation for FOWT dynamic analysis. Fig. 1 illustrates the modules for the coupled simulation and the coupling approach through the transfer of various variables. To simplify the problem, the tower and blades are assumed rigid here, and the floating platform can generally be modelled as a rigid body moving with six degrees of freedom (DOFs). Additionally, we also assume that each component of the FOWT is rigidly connected to each other. Thus, the equation of motion for the FOWT system can be simplified as:

$$M_{ij}\ddot{\xi}_j = F_i^{\text{hydro}} + F_i^{\text{moor}} + F_i^{\text{aero}} - mg\delta_{i3} \tag{1}$$

where $M_{ij}$ is the 6×6 mass matrix; $\xi_j$, $\dot{\xi}_j$ and $\ddot{\xi}_j$ are the motion displacement, velocity and acceleration in the *j*th DOF respectively; $F_i^{\text{hydro}}$ represents the hydrodynamic forces/moments on the platform, described in Section 2.1; $F_i^{\text{moor}}$ denotes the resultant forces/moments on the platform from all mooring lines; $F_i^{\text{aero}}$ denotes the aerodynamic forces on the rotor and corresponding moments about the FOWT system's centre of mass (CoM), and the aerodynamic model is introduced in Section 2.2. Note that only the aerodynamic loads on the blades are considered, while the aerodynamics of the tower, hub and nacelle are currently ignored; $-mg\delta_{i3}$ represents the gravity, where $m$ denotes the mass, $g$ the gravitational acceleration and $\delta_{ij}$ the Kronecker delta.

Another focus of this study is the incorporation of a control system for actively controlling the collective blade pitch and generator torque to adapt to the changes in relative wind speed and ensure smooth rotor operation. The baseline and modified control strategies are detailed in Section 2.3. A brief description of the turbulent wind simulation by TurbSim is given in Section 2.4.

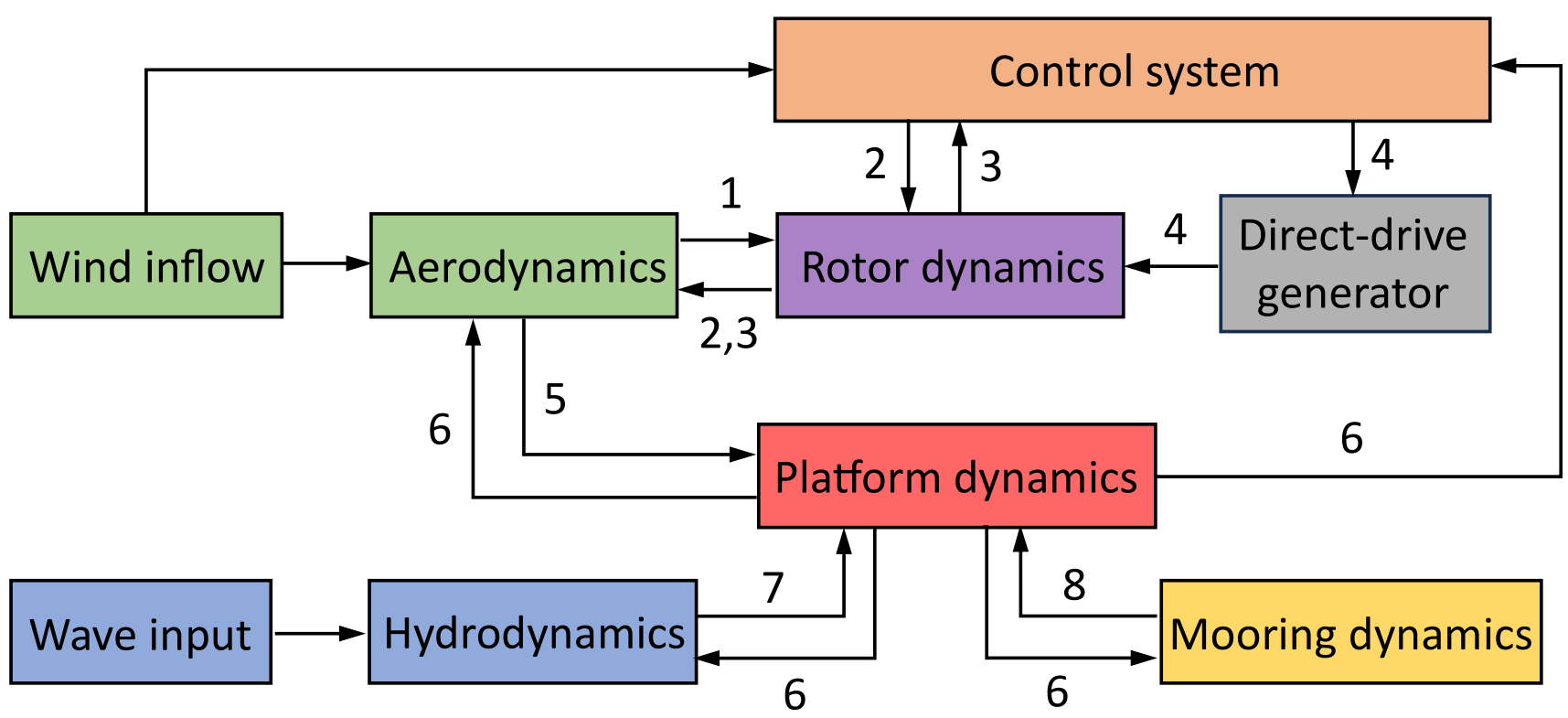

**Fig. 1.** Diagram of coupled aero-hydro-mooring-servo simulation for FOWT dynamics.

## *2.1. Hydrodynamic and mooring models*

The hydrodynamic model OREGEN_X (validated in Zhang et al., 2025) is applied here, mainly based on the linear Cummins method (Cummins, 1962), following Stansby et al. (2019), now with a nonlinear extension of wave excitation forces including instantaneous platform position and nonlinear drag forces. The hydrodynamic forces can be given as:

$$F_i^{\text{hydro}} = F_i^{\text{hs}} + F_i^{\text{exci}} + F_i^{\text{rad}} + F_i^{\text{drag}} \tag{2}$$

where $F_i^{\text{hs}}$ represents the hydrostatic forces, including the hydrostatic restoring forces and buoyancy; $F_i^{\text{exci}}$ denotes the wave excitation forces; $F_i^{\text{rad}}$ represents the radiation forces from the added mass and radiation damping; $F_i^{\text{drag}}$ represents the viscous drag forces by summing drag forces on all platform components (4 columns and 3 beams in this study) through a Morison drag term using appropriate drag coefficients.

This study employs the in-house boundary element method solver OREGEN-BEM (Li and Stansby,

2023) to compute frequency-dependent hydrodynamic coefficients for time-domain hydrodynamic force calculations. For the mooring modelling, both quasi-static theory (Jonkman, 2009) and the dynamic lumped-mass solver MoorDyn (Hall and Goupee, 2015) have been implemented and coupled with OREGEN_X. The details and verification of the coupled hydro-mooring model can be seen in Zhang et al. (2025). Previous work has indicated that mooring model choice contributes marginally to platform motion for chain moorings, though it affects the mooring tension to a large extent particularly for fatigue analysis (see Zhang et al., 2025). The efficient quasi-static mooring model is adopted here, aligning with the focus of this study on platform responses.

### *2.2. BEMT aerodynamic model*

The aerodynamic modelling of wind turbine blades is based on the classical blade element momentum theory (BEMT) with some semi-empirical models to capture unsteady and viscous effects (Hansen, 2008; Apsley and Stansby, 2020). These semi-empirical models include tip- and hub-loss corrections to take into account the vortex shedding at these locations, and the modified Glauert correction (Chapman, 2013; Apsley and Stansby, 2020) when the axial induction factor $a > 0.4$. This in-house aerodynamic solver is referred to as OREGEN_BEMT hereafter. This aerodynamic model has been extended to deal with turbulent wind fields by dividing the rotor plane into multiple azimuthal sectors (azimuthal BEMT). The effects of platform motion on the effective wind speed experienced by the blades and the orientation of rotor plane are also incorporated in OREGEN_BEMT for FOWTs.

Fig. 2 shows the sketch of local velocity and force components for a blade element (2-D aerofoil). The blade element's relative velocity $V_{\text{rel}}$ is the vector sum of the axial velocity $\left(V_{\text{in}} - \tilde{V}_a\right)(1-a)$ and the tangential velocity $\left(\Omega r + \tilde{V}_t\right)(1+a')$ at the rotor plane. The local flow angle $\varphi$ is given by:

$$\tan\varphi = \frac{(V_{\text{in}} - \tilde{V}_a)(1-a)}{(\Omega r + \tilde{V}_t)(1+a')} \tag{3}$$

where $V_{\text{in}}$ is the inflow wind speed, $\Omega$ is the rotor's rotational speed, $r$ is the radial position of the blade element along the span, $\tilde{V}_a$ and $\tilde{V}_t$ denote the additional axial and tangential velocities induced by platform motions, $a$ and $a'$ denote the axial and tangential induction factors. $\theta$ is the local pitch angle of the blade element, which is a sum of the blade pitch angle $\beta$ regulated by the pitch controller and the blade twist $\theta_{\text{twist}}$ (i.e., $\theta = \theta_{\text{twist}} + \beta$). Then the local angle of attack $\alpha$ derives from:

$$\alpha = \varphi - \theta \tag{4}$$

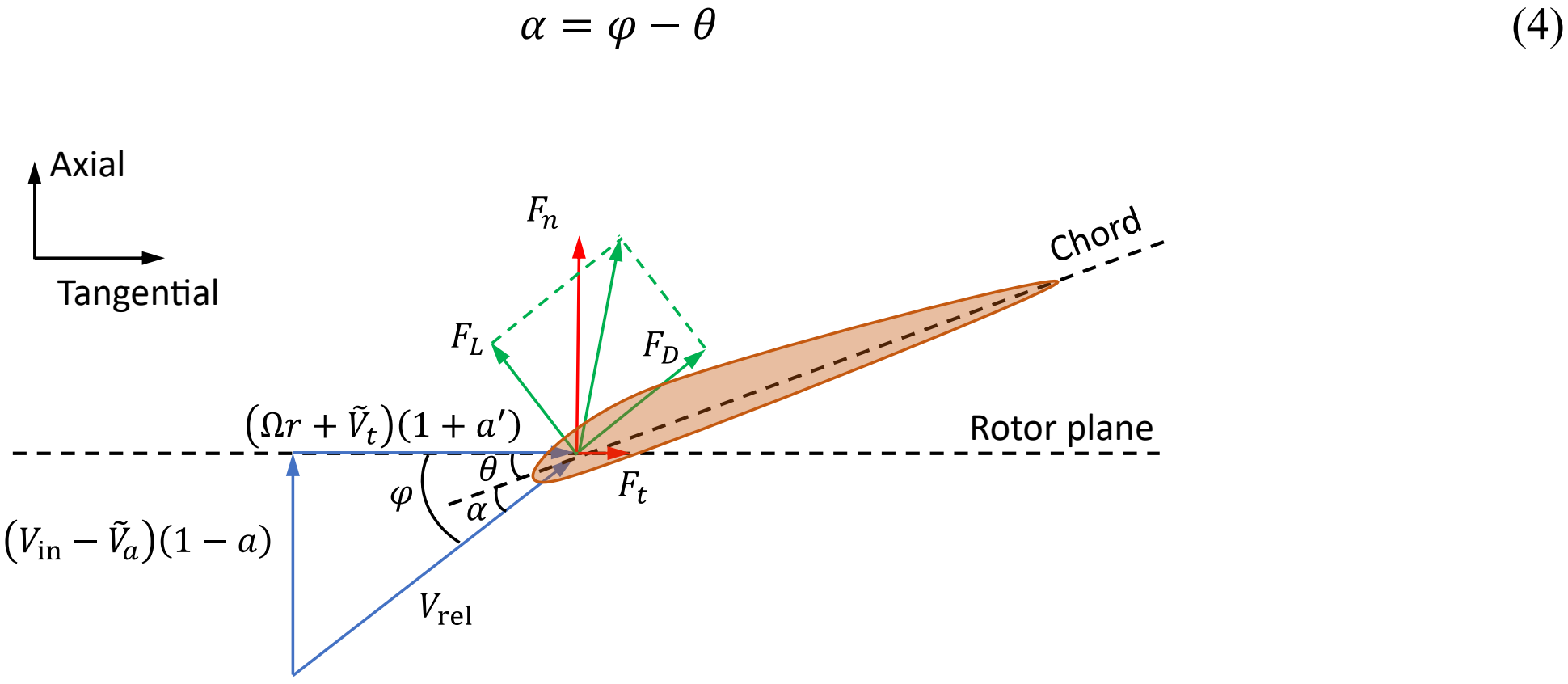

**Fig. 2.** Sketch of local velocity and force components for a blade element (2-D aerofoil).

Following the local flow conditions shown in Fig. 2, the aerodynamic forces on the blade element can be computed. The lift $F_L$ and drag $F_D$ are projected into the directions that are normal and tangential to the rotor plane, i.e., $F_n$ and $F_t$. Then the thrust on an annular sector of width $dr$ can be derived as:

$$dF_n = \frac{1}{2N_\theta} \rho_a V_{\mathrm{rel}}^2 B c C_n dr \tag{5}$$

and the torque from blade elements in the annular sector is written as:

$$dM = r dF_t = \frac{1}{2N_\theta} \rho_a V_{\mathrm{rel}}^2 B c C_t r dr \tag{6}$$

where $N_\theta$ is the number of azimuthal sectors, $\rho_a$ is the air density, $c$ is the chord length, $B$ is the number of blades, $C_n$ and $C_t$ denote the normal and tangential coefficients, the expressions of which are written as:

$$C_n = C_l \cos\varphi + C_d \sin\varphi \tag{7}$$

$$C_t = C_l \sin\varphi - C_d \cos\varphi \tag{8}$$

where $C_l$ and $C_d$ denote the lift and drag coefficients obtained from the aerofoil look-up tables as functions of the local angle of attack.

Using momentum theory, the expressions of local thrust $dF_n$ and torque $dM$ on an annular sector can also be found. Integrating blade element and momentum theories, we can establish an iterative process to determine $a$ and $a'$ at each blade element and then compute the total aerodynamic loads by integration along the blade span. The convergence criterion for the iteration of $a$ and $a'$ is that their absolute differences between successive iterations are less than $10^{-6}$. More details can be found in Apsley and Stansby (2020), and a flow chart is given in Fig. 3 to briefly illustrate the numerical algorithm of the BEMT model.

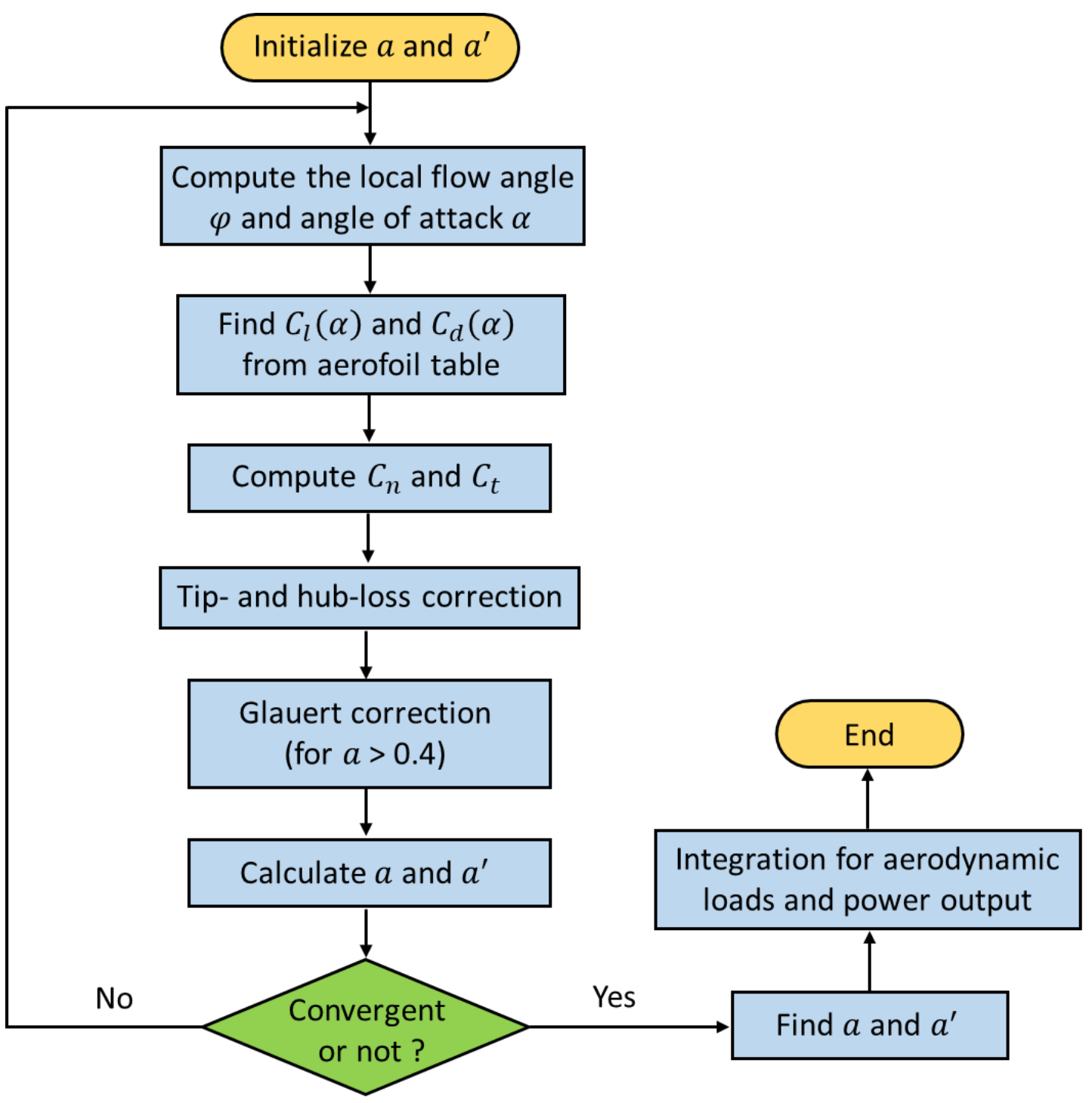

**Fig. 3.** Flow chart of the numerical algorithm for the BEMT aerodynamic model.

### 2.3. *Rotor control*

#### 2.3.1 *Baseline controller*

The baseline controller is designed for bottom-fixed or land-based wind turbines without accounting for floating effects. A first-order direct-drive wind turbine model follows Pao and Johnson (2011):

$$I_r\dot{\Omega} = \tau_a - \tau_g \tag{9}$$

where $\Omega$ is the rotor speed, $I_r$ is the rotor inertia, $\tau_a$ is the aerodynamic torque, which is a nonlinear function of rotor variables ($\Omega$, $V$ and $\beta$) calculated by an aerodynamic model, and $\tau_g$ is the generator torque.

Linearising Eq. (9) at a steady-state operating point yields:

$$I_r\dot{\tilde{\Omega}} = \frac{\partial\tau_a}{\partial\Omega}\tilde{\Omega} + \frac{\partial\tau_a}{\partial\beta}\tilde{\beta} + \frac{\partial\tau_a}{\partial V}\tilde{V} - \tilde{\tau}_g \tag{10}$$

where the notation $\tilde{x}$ denotes a perturbation from the steady-state operating point $\bar{x}$, so $x = \tilde{x} + \bar{x}$, in which *x* represents an arbitrary variable. The equation can then be rewritten as:

$$\dot{\tilde{\Omega}} = A\tilde{\Omega} + B^{(\beta)}\tilde{\beta} + B^{(V)}\tilde{V} + B^{(\tau)}\tilde{\tau}_g \tag{11}$$

where $A = \frac{1}{I_r}\frac{\partial\tau_a}{\partial\Omega}$, $B^{(\beta)} = \frac{1}{I_r}\frac{\partial\tau_a}{\partial\beta}$, $B^{(V)} = \frac{1}{I_r}\frac{\partial\tau_a}{\partial V}$ and $B^{(\tau)} = -\frac{1}{I_r}$. These coefficients can be numerically estimated at each steady-state operating point ($\bar{\Omega}$, $\bar{V}$ and $\bar{\beta}$) by the aerodynamic solver OREGEN_BEMT. The steady-state $\bar{\Omega}$, $\bar{\beta}$ and $\bar{\tau}_g$ are derived with respect to $\bar{V}$ at predefined wind speed conditions.

A reference tracking PI controller operates in above/below-rated regions with the general form:

$$y = k_p u + k_i \int_0^t u dt \tag{12}$$

where *u* is the controller input (error signal), *y* is the controller output (control signal), $k_p$ and $k_i$ denote the proportional and integral gains, respectively.

In above-rated operation, the collective blade pitch is adjusted by a PI controller to maintain rated rotor speed, while the generator torque is kept constant at its rated value ($\tilde{\tau}_g = 0$). The blade pitch controller has the following inputs and outputs:

$$u = \tilde{\Omega},\ y = \tilde{\beta} \tag{13}$$

Note that the wind speed disturbance $\tilde{V}$ is zeroed for the controller tuning. Thus, for the blade pitch controller tuning, Eq. (11) can be given by:

$$\dot{\tilde{\Omega}} = A\tilde{\Omega} + B^{(\beta)}k_p^{(\beta)}\tilde{\Omega} + B^{(\beta)}k_i^{(\beta)} \int_0^t \tilde{\Omega} dt \tag{14}$$

where $k_p^{(\beta)}$ and $k_i^{(\beta)}$ are the PI gains for the blade pitch controller. We also define the rotor speed integral error as $\tilde{\vartheta} = \int_0^t \tilde{\Omega} dt$, such that $\dot{\tilde{\vartheta}} = \tilde{\Omega}$ and $\ddot{\tilde{\vartheta}} = \dot{\tilde{\Omega}}$. The closed-loop dynamics are then modelled by a second-order differential equation:

$$\ddot{\tilde{\vartheta}} - \left(A + B^{(\beta)} k_p^{(\beta)}\right) \dot{\tilde{\vartheta}} - B^{(\beta)} k_i^{(\beta)} \tilde{\vartheta} = 0 \tag{15}$$

The desired closed-loop transient response can be achieved by specifying a desired natural frequency $\omega_{\text{des}}$ and damping ratio $\zeta_{\text{des}}$. Increasing $\omega_{\text{des}}$ reduces the response time of the rotor, while increasing $\zeta_{\text{des}}$ reduces the amount of oscillation in the response. $\omega_{\text{des}}$ is generally selected to be relatively large to ensure sufficient control bandwidth for rotor speed regulation, whereas a smaller $\omega_{\text{des}}$ is often adopted as a detuning strategy to mitigate the negative damping effect in FOWTs. Then the PI gains for the blade pitch controller can be analytically derived (Abbas et al., 2022; Stockhouse et al., 2023) as:

$$k_p^{(\beta)} = -\left(\frac{2\zeta_{\text{des}}\omega_{\text{des}} + A}{B^{(\beta)}}\right) \tag{16}$$

$$k_i^{(\beta)} = -\frac{\omega_{\text{des}}^2}{B^{(\beta)}} \tag{17}$$

For below-rated wind speeds, the PI generator torque controller aims to maximize power capture by tracking optimal TSR with the constraint on minimum rotor speed, while the blade pitch angle is fixed at its nominal value ($\bar{\beta} = 0$). The generator torque controller has the following inputs and outputs:

$$u = \tilde{\Omega},\ y = \tilde{\tau}_g \tag{18}$$

Similarly, the PI gains for the generator torque controller are defined as:

$$k_p^{(\tau)} = -\left(\frac{2\zeta_{\text{des}}\omega_{\text{des}} + A}{B^{(\tau)}}\right) \tag{19}$$

$$k_i^{(\tau)} = -\frac{\omega_{\text{des}}^2}{B^{(\tau)}} \tag{20}$$

Therefore, once the tuning parameters ($\omega_{\text{des}}$ and $\zeta_{\text{des}}$) are defined, we can compute the PI gains for the controllers to satisfy the desired behaviour. For control designers, specifying closed-loop performance through $\omega_{\text{des}}$ and $\zeta_{\text{des}}$ is generally more intuitive than directly tuning $k_p$ and $k_i$. The PI controller gains are scheduled by $\bar{V}$ at predefined steady-state operating points and interpolated afterwards according to the current operating point.

Note that though the present baseline controller is theoretically the same as that of ROSCO (Abbas et al., 2022), the numerical implementation is slightly different. The PI gains for the blade pitch controller are scheduled by $\bar{\beta}$ in ROSCO for its simplicity due to the monotonic relationship between $\bar{V}$ and $\bar{\beta}$ in this region. The in-house aerodynamic solver OREGEN_BEMT provides the aerodynamic sensitivities in Eq. (11) instead of using AeroDyn module of OpenFAST. We also emphasize that for FOWT the nominal steady-state operating point here is chosen based on the relative wind speed incorporating the hub fore–aft velocity instead of the inflow wind speed.

### 2.3.2 *Platform motion feedback*

A common control characteristic for FOWTs involves implementing an auxiliary control loop that incorporates feedback from the rotor fore-aft velocity, usually measured at the nacelle. This is referred to as parallel compensation (van der Veen et al., 2012), and has been shown to be effective when implemented with blade pitch or generator torque as the compensating actuator (e.g., Fischer, 2013; Yu et al., 2018; Abbas et al., 2022; Stockhouse et al., 2023).

The ROSCO (Abbas et al., 2022) implements compensation on the blade pitch and uses a constant proportional gain for the feedback term. However, this study implements gain-scheduled compensations for both blade pitch and generator torque. Although some studies suggest using a platform pitch rate feedback signal (e.g., Yu et al., 2018; Stockhouse et al., 2022; Capaldo and Mella, 2023), the hub/nacelle fore–aft velocity signal is used here so that both platform surge and pitch motions are incorporated, which is also usually more consistent with the physical measurements of nacelle acceleration in practice.

The compensations for the blade pitch ($\tilde{\beta}_c$) and generator torque ($\tilde{\tau}_{g,c}$) are given by a proportional term as:

$$\tilde{\beta}_c = k_\beta \dot{x}_{\mathrm{hub}} \tag{21}$$

$$\tilde{\tau}_{g,c} = k_\tau \dot{x}_{\mathrm{hub}} \tag{22}$$

where $k_\beta$ and $k_\tau$ are the proportional gains for blade pitch and generator torque compensations respectively, and $\dot{x}_{\mathrm{hub}}$ is the hub fore-aft velocity, combining platform surge velocity $\dot{\xi}_1$ and pitch-induced linear velocity $h_{\mathrm{t}}\dot{\xi}_5$ under the small-angle approximation:

$$\dot{x}_{\mathrm{hub}} = \dot{\xi}_1 + h_{\mathrm{t}}\dot{\xi}_5 \tag{23}$$

where $h_{\mathrm{t}}$ is the hub height above the FOWT's CoM.

Once $k_\beta$ and $k_\tau$ are determined, $\tilde{\beta}_c$ and $\tilde{\tau}_{g,c}$ are superimposed on the baseline PI controller outputs using Eqs. (21) and (22), respectively. This leads to $\tilde{\beta} = \tilde{\beta}_{PI} + \tilde{\beta}_c$ and $\tilde{\tau}_g = \tilde{\tau}_{g,PI} + \tilde{\tau}_{g,c}$, where $\tilde{\beta}_{PI}$ and $\tilde{\tau}_{g,PI}$ denote blade pitch and generator torque control signals generated by the PI controller described in Section 2.3.1. In this study, the choices of $k_\beta$ and $k_\tau$ are novel compared to other references (e.g., Abbas et al., 2022; Stockhouse et al., 2022, 2024), and are automatically tuned using the methodology introduced below.

Similar to the linearisation of $\tilde{\tau}_a$ in Eq. (10), the thrust variation $\tilde{F}_a$ can be linearised at a given operating point as:

$$\tilde{F}_a = \frac{\partial F_a}{\partial \Omega}\tilde{\Omega} + \frac{\partial F_a}{\partial \beta}\tilde{\beta} + \frac{\partial F_a}{\partial V}\tilde{V} \tag{24}$$

Considering the hub fore–aft velocity due to platform motions, the rotor effective wind speed is modified as the relative wind speed at the rotor. Assuming $\dot{\bar{x}}_{\mathrm{hub}} = 0$ ($\dot{\tilde{x}}_{\mathrm{hub}} = \dot{x}_{\mathrm{hub}}$), the relative wind speed disturbance $\tilde{V}$ is given by:

$$\tilde{V} = \tilde{V}_w - \dot{x}_{\mathrm{hub}} \tag{25}$$

where $\tilde{V}_w$ is the free-stream (inflow) wind perturbation. Accounting for platform motions, the thrust in Eq. (24) and the aerodynamic torque in Eq. (10) employ this relative wind speed.

Including the blade pitch compensation ($\tilde{\beta}_c$), now the blade pitch control signal becomes:

$$\tilde{\beta} = k_p^{(\beta)}\tilde{\Omega} + k_i^{(\beta)} \int_0^t \tilde{\Omega} dt + k_\beta \dot{x}_{\text{hub}} \tag{26}$$

Substituting Eq. (25) and Eq. (26) into Eq. (24), then we can get the following equation for the additional thrust perturbation arising from the hub velocity:

$$\Delta F_a = \left(\frac{\partial F_a}{\partial \beta} k_\beta - \frac{dF_a}{dV}\right) \dot{x}_{\text{hub}} \tag{27}$$

where $\frac{dF_a}{dV}$ represents the ideal closed-loop thrust sensitivity to wind speed (Jonkman, 2008), as $\Omega$ and $\beta$ are related to $V$. The linearised aerodynamic damping coefficient $C_{\text{aero}}$ related to $\dot{x}_{\text{hub}}$ can be estimated as:

$$C_{\text{aero}} = \frac{dF_a}{dV} - k_\beta \frac{\partial F_a}{\partial \beta} \tag{28}$$

Note that the first term of $C_{\text{aero}}$, i.e., $\frac{dF_a}{dV}$, may become negative in above-rated wind conditions using the baseline controller, and thus $C_{\text{aero}}$ would be negative if the $k_\beta$ term is neglected. To avoid potential platform motion instability caused by the negative damping effect, a proper $k_\beta$ should be chosen to ensure that $C_{\text{aero}}$ is non-negative. If we set $C_{\text{aero}} = 0$, we can derive the expression of $k_\beta$ as:

$$k_\beta = \frac{dF_a}{dV} \left(\frac{\partial F_a}{\partial \beta}\right)^{-1} \tag{29}$$

Note that here $k_\beta$ adapts to the wind speed with gain scheduling instead of keeping a constant gain.

For the rotor dynamics calculated by Eq. (10), the additional torque perturbation arising from the hub velocity due to platform motions can be given by:

$$\Delta \tau = \frac{\partial \tau_a}{\partial \beta} k_\beta \dot{x}_{\text{hub}} - \frac{d\tau_a}{dV} \dot{x}_{\text{hub}} - k_\tau \dot{x}_{\text{hub}} \tag{30}$$

Similarly, $\frac{d\tau_a}{dV}$ represents the closed-loop torque sensitivity to wind speed using the baseline controller. To eliminate the influence of hub velocity on rotor dynamics, we specify $k_\tau$ as:

$$k_\tau = -\frac{d\tau_a}{dV} + k_\beta \frac{\partial \tau_a}{\partial \beta} \tag{31}$$

In this way, we expect to mitigate the negative damping effect on platform motions and decouple rotor dynamics from platform dynamics to get better rotor operating performance (i.e., smaller rotor acceleration).

The control modules used in this study are summarized concisely in the block diagram shown in Fig. 4. These control modules ensure smooth rotor operation across all operational regions. For turbulent wind, we may use the low-pass filtered hub-height or rotor-disk-averaged wind speed as the wind input to determine the steady-state operating point for control purposes, and a first-order low-pass filter is also adopted to filter the nacelle/hub velocity signal to avoid spikes in platform motion feedback.

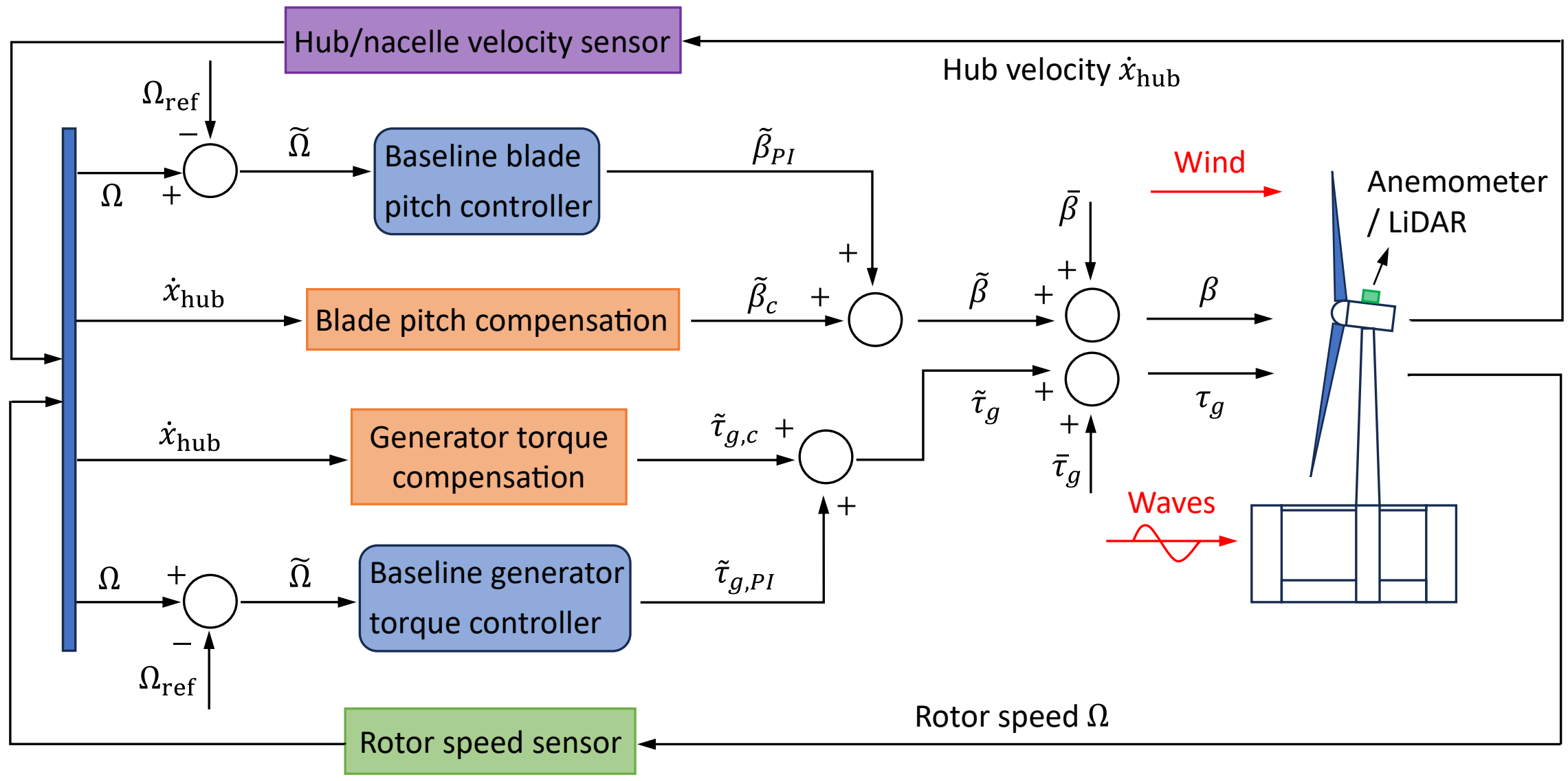

**Fig. 4.** Block diagram of the control modules used in this study. ($\Omega_{\text{ref}}$ represents the reference rotor speed, $\tilde{\Omega} = \Omega - \Omega_{\text{ref}}$ represents the rotor speed error, $\tilde{\beta}_{PI}$ and $\tilde{\tau}_{g,PI}$ denote blade pitch and generator torque control signals generated by the PI controller, $\tilde{\beta}_c$ and $\tilde{\tau}_{g,c}$ denote the blade pitch and generator torque compensations from platform motion feedback only for FOWTs, $\bar{\beta}$ and $\bar{\tau}_g$ denote the steady-state blade pitch and generator torque related to effective wind speed)

### *2.4. Turbulent wind simulation*

In this study, a synthetic turbulent shear wind field is generated using TurbSim (Jonkman, 2014) based on the IEC Kaimal spectra, once the mean hub-height wind speed $\bar{u}_{\text{hub}}$ and turbulence intensity TI (or a standard IEC turbulence category) are specified. The wind shear exponent uses a reference vale of 0.14 for offshore conditions.

Using the wind field data from TurbSim, we can obtain the rotor-disk-averaged wind speed, which is used as the rotor effective wind speed for control in turbulent wind cases. This is calculated using plane polar coordinates as:

$$\bar{u}_{\text{RA}} = \frac{1}{S}\iint u(y,z)dS = \frac{1}{S}\int_0^{2\pi}\int_{R_{\text{hub}}}^{R_{\text{rotor}}} u(r\cos\theta, r\sin\theta + H_{\text{hub}})rdrd\theta \tag{32}$$

where $S = \pi\left(R_{\text{rotor}}^2 - R_{\text{hub}}^2\right)$, $R_{\text{rotor}}$ and $R_{\text{hub}}$ denote the radius of the rotor and hub, $H_{\text{hub}}$ is the hub height above still water level (SWL). If the rotor disk is divided into $N_r$ radial intervals and $N_\theta$ angular intervals, the discrete form of Eq. (32) can be written as:

$$\bar{u}_{\text{RA}} = \frac{1}{S}\sum_{i=1}^{N_r}\sum_{j=1}^{N_\theta} u_{ij}\,\Delta S_{ij} \tag{33}$$

where $\Delta S_{ij} = r_i \Delta r \Delta\theta$, in which $r_i = R_{\text{hub}} + (i - 0.5)\Delta r$, $\Delta r = (R_{\text{rotor}} - R_{\text{hub}})/N_r$, and $\Delta\theta = 2\pi/N_\theta$. The wind speed $u_{ij}$ is at a point $(r_i, \theta_j)$ in polar coordinates, and its corresponding Cartesian coordinates are $(r_i\cos\theta_j, r_i\sin\theta_j + H_{\text{hub}})$. Thus, the value of $u_{ij}$ can be obtained by a 2-D linear interpolation of the wind field in rectangular grids from TurbSim, as shown in Fig. 5. The inflow wind speed on the blade nodes used in the thrust calculation is also obtained in the same way.

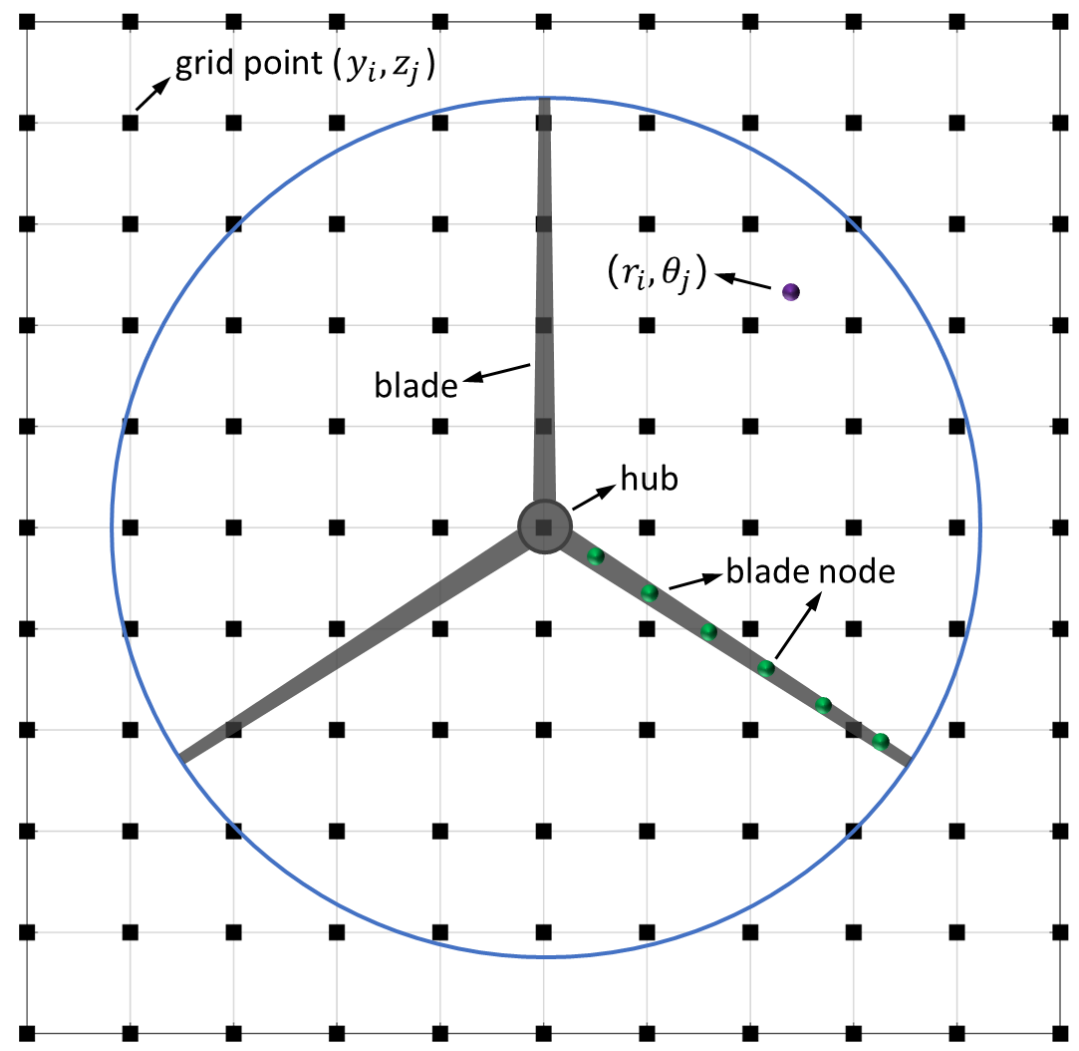

**Fig. 5.** Sketch of the rectangular grid used in TurbSim and interpolation of wind speed at any point on the rotor plane including blade nodes. The blue circle denotes the rotor diameter.

## 3. FOWT description and numerical setup

This study models a FOWT system based on the IEA Wind 15 MW reference wind turbine (Gaertner et al., 2020) mounted on the Volturn platform (Allen et al., 2020), as can be seen in Fig. 6. The Class IB direct-drive turbine features a 240 m rotor diameter and 150 m hub height above SWL. The Volturn platform utilizes a central tower-supporting column connected via three submerged rectangular pontoons to equally spaced outer columns. Design specifications include a 20 m draft, 15 m freeboard, and 200 m operational water depth.

Table 1 lists the key properties of the 15 MW reference FOWT system including the rotor nacelle assembly (RNA), wind tower and platform. Note the moments of inertia and CoM of the FOWT system may be slightly different from those in Allen et al. (2020), since the report only gives the platform's moments of inertia and CoM. These values in Table 1 correspond to the 1:70 scaled experiment (Ransley et al., 2022) against which the proposed hydro-mooring model was validated (see Zhang et al., 2025), and they are further confirmed by the natural frequencies, which closely match the reported values. For the mooring system, three chain catenary moorings are used with the same length of 850 m. The dry line density of each mooring line is 685 kg/m with the volume-equivalent diameter of 0.333 m, and the extensional stiffness (EA) is 3270 MN. More details about the wind turbine and platform properties can be found in Gaertner et al. (2020) and Allen et al. (2020). The detailed blade properties, such as the spanwise position, chord, twist and aerodynamic characteristics (i.e., lift and drag coefficients) of the aerofoils used along the blade can be obtained by look-up tables on GitHub: https://github.com/IEAWindSystems/IEA-15-240-RWT/blob/master/Documentation/IEA-15-240-RWT_tabular.xlsx (last access: 10 December 2024).

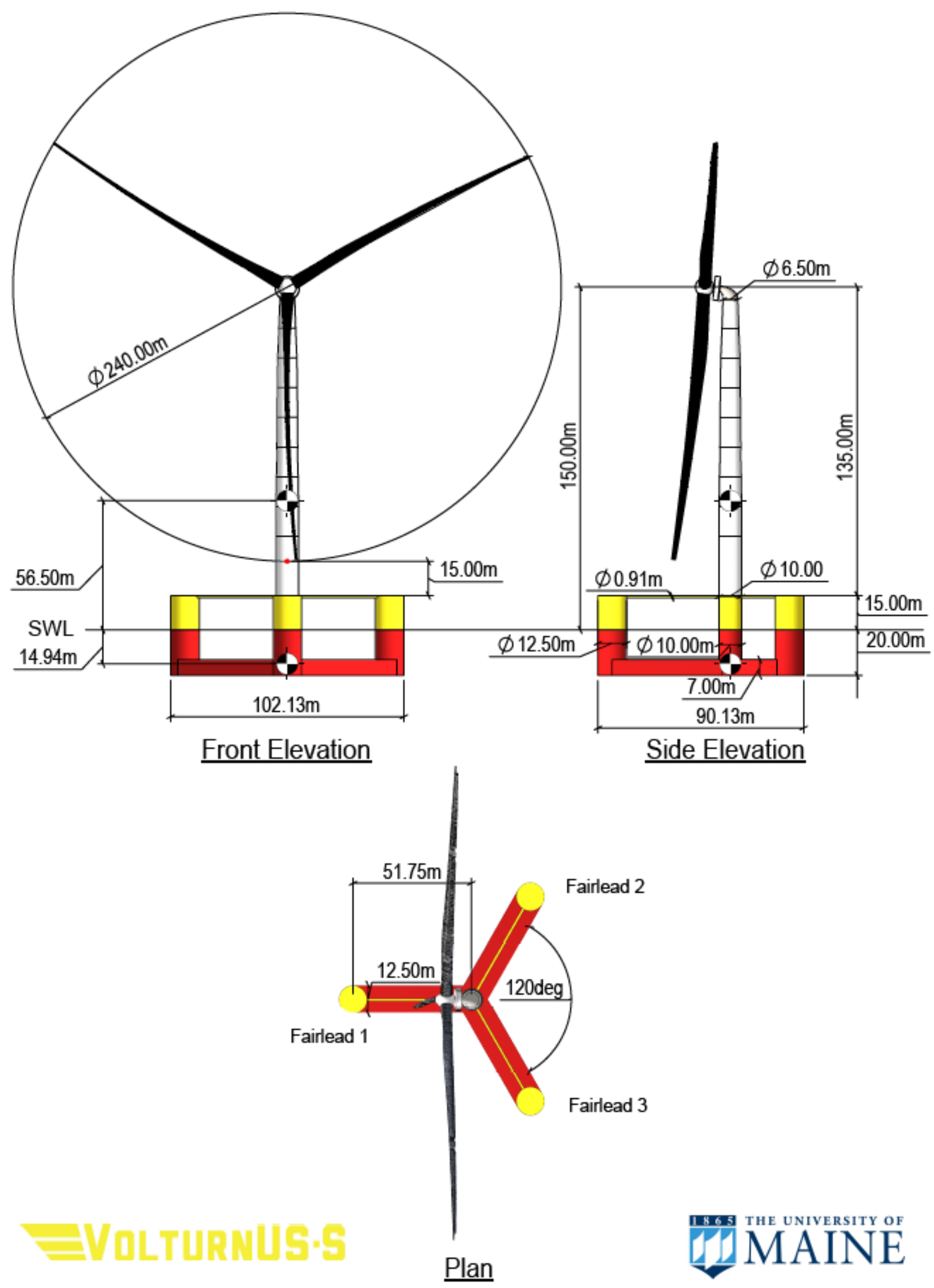

**Fig. 6.** Diagram of the 15 MW reference FOWT using Volturn platform with dimensions, from Allen et al. (2020) with permission.

**Table 1.** Key properties of the 15 MW reference FOWT system.

| Parameter | Value |
|---|---|
| Total system mass | 2.0093E+07 kg |
| Roll and pitch inertia about CoM | 4.4841E+10 kg-m$^2$ |
| CoM position from SWL | -1.67 m |
| Centre of buoyancy from SWL | -13.63 m |
| Displacement | 20206 m$^3$ |

For aerodynamic modelling, the rotor disk is divided into 50 radial intervals and 36 angular intervals ($N_r = 50$, $N_\theta = 36$) in this study, which is found to yield nearly interval-independent results. For turbulent wind simulations by TurbSim, a mesh of 21 × 21 grid points covering a rectangular surface of 260 × 260 m is employed. The time step used in TurbSim is the same as that in the aerodynamic model, which is set to 0.05 s for the fixed rotor and equal to the time step used in the hydrodynamic model for the FOWT, i.e., 1/200 of the wave period in this study.

For the controller setup and tuning, $\omega_{\text{des}}$ and $\zeta_{\text{des}}$ are the only parameters that need to be defined. As the steady-state operating point is determined based on the relative wind speed for FOWTs in this paper, the actual operating point under the baseline controller remains close to the steady-state one. Therefore, relatively small PI gains are sufficient to maintain stability without introducing excessive control action. Here, $\omega_{\text{des}} = 0.04$ rad/s and $\zeta_{\text{des}} = 1.0$ are chosen for the blade pitch controller after some tests. To avoid unrealistically high gains near rated operation, $k_p^{(\beta)}$ and $k_i^{(\beta)}$ are approximated via linear fits at near-rated operating point. For the generator torque controller, we set $\omega_{\text{des}} = 0.08$ rad/s and $\zeta_{\text{des}} = 1.0$.

Considering the physical limit of the blade pitch and generator torque controller, there are some constraints on their minimum/maximum magnitudes and maximum rotation speed. The minimum and maximum blade pitch angles are set to 0° and 25° within the operational wind speeds, and the maximum blade pitch rate (in absolute value) is set to 2°/s unless specifically mentioned. The generator torque is constrained to its rated value (around 20 MN-m) with a maximum slew rate of $\pm$4.5 MN-m/s.

## 4. Results and discussions

We first run simulations for the bottom-fixed wind turbine. The steady-state performance of the turbine under steady uniform wind is validated and some turbulent wind tests are investigated. Afterwards, the simulations for the FOWT with different control strategies and different types of wind/wave inputs are implemented. We demonstrate negative damping phenomenon with baseline control and evaluate control strategy and turbulent wind impacts on platform response and rotor performance.

### *4.1. Bottom-fixed wind turbine*

#### *4.1.1. Steady wind performance*

Fig. 7 shows the steady-state operating points for the bottom-fixed wind turbine across the operational wind speed range (3–25 m/s). The controller's functionality is characterized by three regions consistent with the definition in NREL report (Gaertner et al., 2020) and Abbas et al. (2022). To help understand and explain, the contour surfaces of power coefficient $C_P$ and thrust coefficient $C_T$ versus blade pitch and TSR are given in Fig. 8. Region 1.5 represents constrained operation between cut-in and rated wind speeds, where the minimum rotor speed limit (5.0 rpm) forces suboptimal TSR. A PI controller on the generator torque is used to maintain the minimum rotor speed, and a minimum blade pitch schedule is implemented to maximize $C_P$, which also leads to a decrease of $C_T$, as indicated by the black dash–dotted line in Fig. 8. Region 2 employs a generator torque controller to maintain optimal TSR (TSR = 9), and the blade pitch angle is fixed at $\beta = 0°$ to achieve $C_P$-maximizing (see the blue dot in Fig. 8). Region 3 represents operation in above-rated wind speeds. The rotor speed is regulated to its rated value of 7.56 rpm via blade pitch control, reducing TSR per the red dashed line in Fig. 8, while the generator torque remains constant.

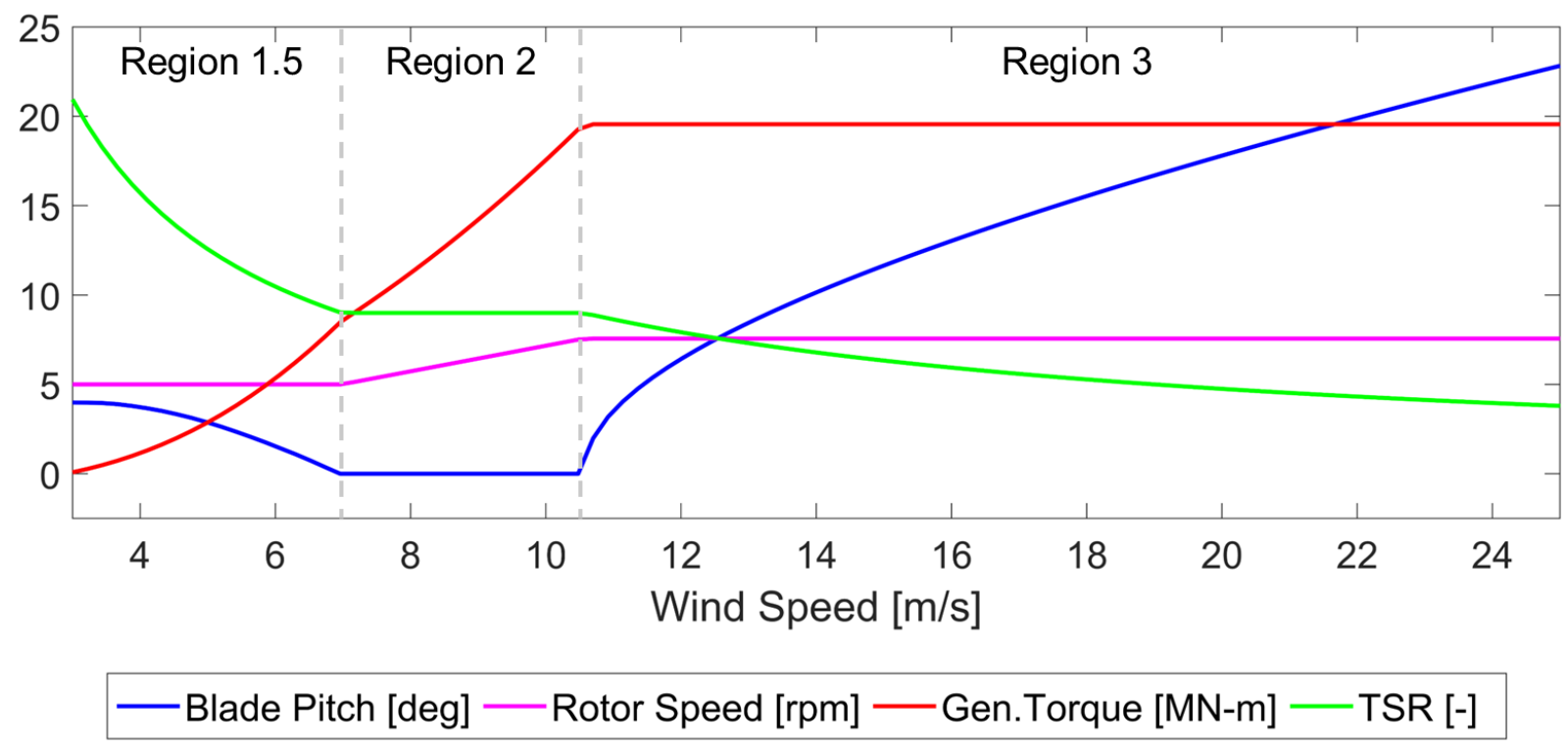

**Fig. 7.** The steady-state operating points for the bottom-fixed IEA 15MW wind turbine.

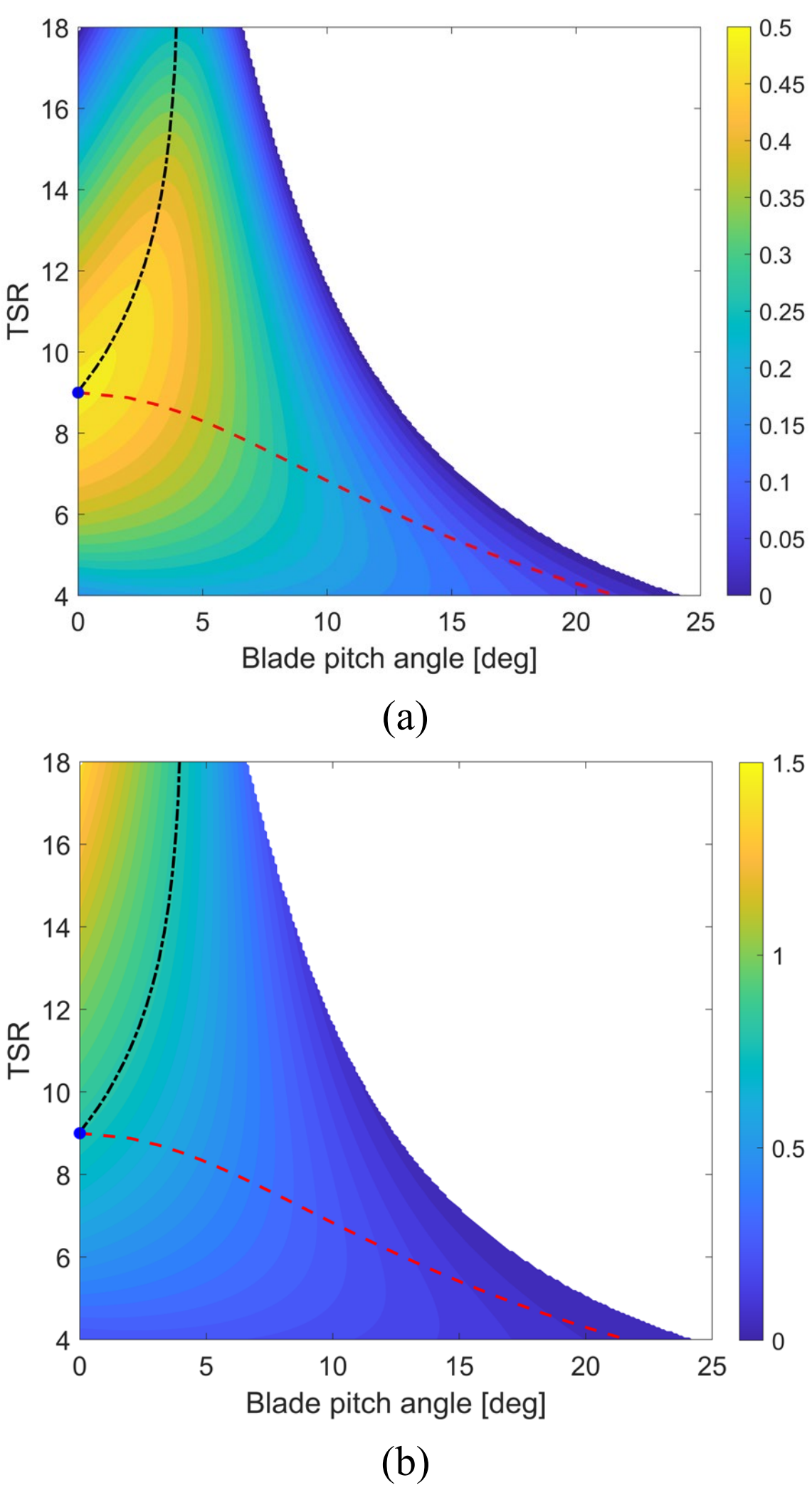

**Fig. 8.** The contour surfaces of power coefficient $C_P$ and thrust coefficient $C_T$ for the bottom-fixed IEA 15MW wind turbine. (a) $C_P$; (b) $C_T$. (The black dash–dotted line denotes the operating points in Region 1.5, the blue dot denotes the operating points in Region 2 and the red dashed line denotes the operating points in Region 3)

To validate the present BEMT model with rotor control, $C_P$ and $C_T$ are compared with those in NREL report (Gaertner et al., 2020) as a function of wind speed, as shown in Fig. 9. Note that the scatter data in the figure are digitized from the report. The present results closely match with those reported by NREL.

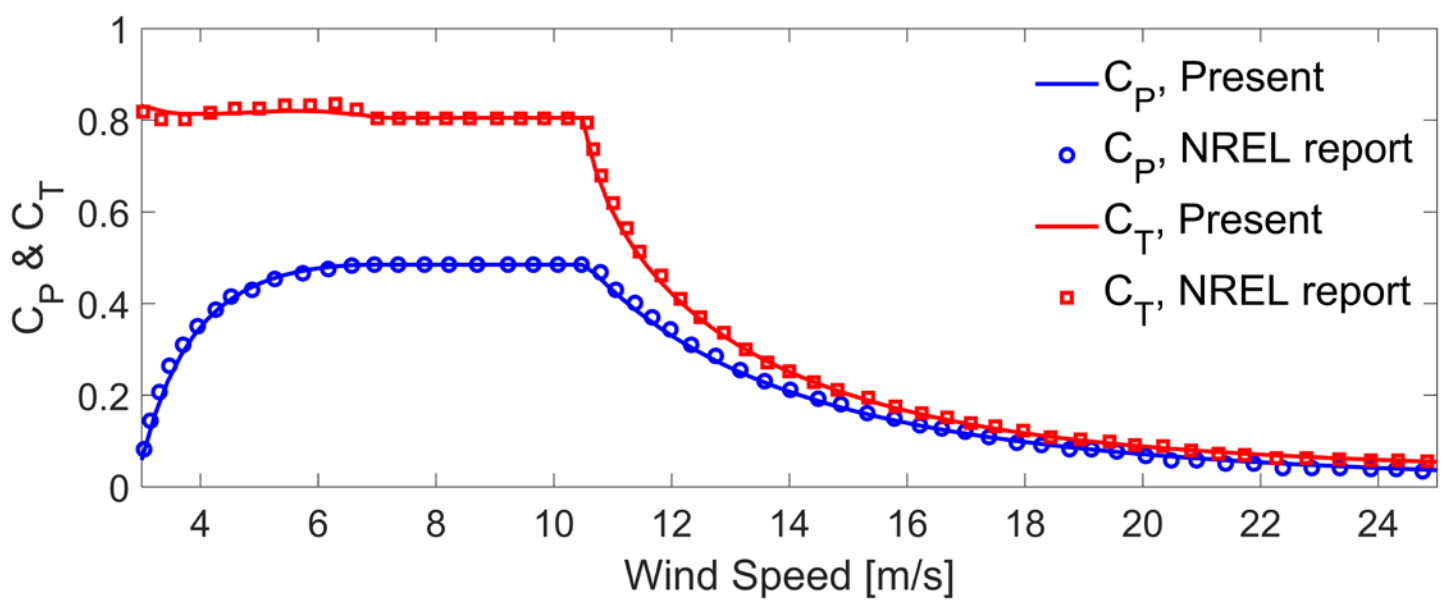

**Fig. 9.** The aerodynamic coefficients $C_P$ and $C_T$ for the bottom-fixed IEA 15MW wind turbine.

#### *4.1.2. Turbulent wind results*

For turbulent wind simulation, the onset wind field is generated by TurbSim with the mean hub-height wind speed ($\bar{u}_{\text{hub}}$) and turbulence intensity (TI) listed in Table 2.

**Table 2.** Mean hub-height wind speed and turbulence intensity for the turbulent wind cases.

| Case ID | 1 | 2 | 3 | 4 | 5 | 6 |
| --- | --- | --- | --- | --- | --- | --- |
| $\bar{u}_{\text{hub}}$ (m/s) | 5 | 8 | 10 | 12 | 15 | 20 |
| TI | 20% | 15% | 15% | 15% | 10% | 10% |

We test two kinds of wind speed inputs as the rotor effective wind speed (REWS) for control purposes. One is the hub-height wind speed (HHwind) corresponding to the wind speed measured by hub-height anemometer commonly in practice. Another is the rotor-disk-averaged wind speed (RAwind), which can be calculated by the wind field data from TurbSim in the simulation (see Section 2.4) and estimated in practice by spinner LiDAR through certain processing and analysis. Generally the spinner LiDAR is more advanced and suitable for the on-site wind speed measurements than the hub-height anemometer, as the hub-height anemometer measurements are highly disturbed by the turbine wake and inadequately characterize the turbulent wind field across a large rotor. The calculation of wind thrust remains the same and is both performed by integrating the aerodynamic forces on blade elements exposed to the turbulent wind field.

Fig. 10 shows an example of 10-min time series results with $\bar{u}_{\text{hub}}$ = 15 m/s and TI = 10%. It is seen from Fig. 10(a) that the fluctuation of RAwind is much smoother than that of HHwind as expected. This is mainly due to that RAwind takes a spatial average over the rotor plane while HHwind is solely from a single point at the hub. The variations of the wind thrust, generator power, rotor speed and blade pitch angle with time using RAwind for control are all smaller than those using HHwind, as shown in Fig. 10(b)-(e).

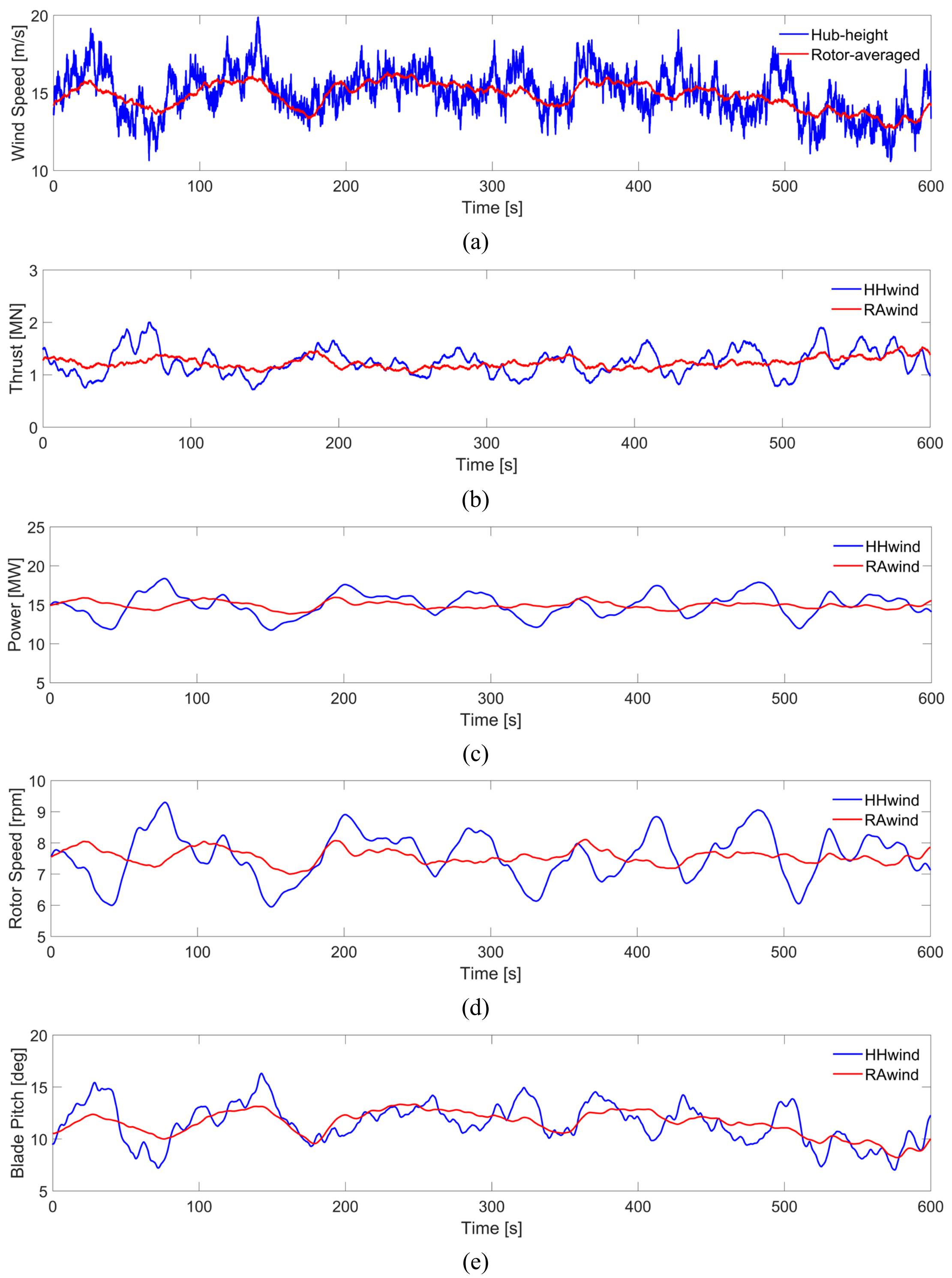

**Fig. 10.** Results for the bottom-fixed IEA 15 MW wind turbine under turbulent shear wind with $\bar{u}_{\text{hub}}$ = 15 m/s and TI = 10%. (a) Wind speed; (b) Wind thrust; (c) Generator power; (d) Rotor speed; (e) Blade pitch.

Fig. 11 shows the turbine performance and operation statistics with 10-min simulation time for the turbulent shear wind cases in Table 2. In the figure, the dots or crosses show the time-averaged values and the error bars denote the standard deviations of the results. The time-averaged values using HHwind and RAwind for control are both quite close to the steady-state results under steady uniform wind with the same $\bar{u}_{\text{hub}}$, except for the wind thrust and blade pitch angle at $\bar{u}_{\text{hub}}$ = 10 m/s due to large variation with wind speed near the rated wind speed (i.e., 10.59 m/s). However, the standard deviations of the wind thrust, generator power and rotor speed results using RAwind for control are significantly less than those using HHwind particularly at above-rated wind speeds. This can mitigate the rotor overspeed and facilitate the smooth turbine operation. As a result, RAwind is proposed as the REWS for control and used in the following turbulent wind simulations for FOWTs.

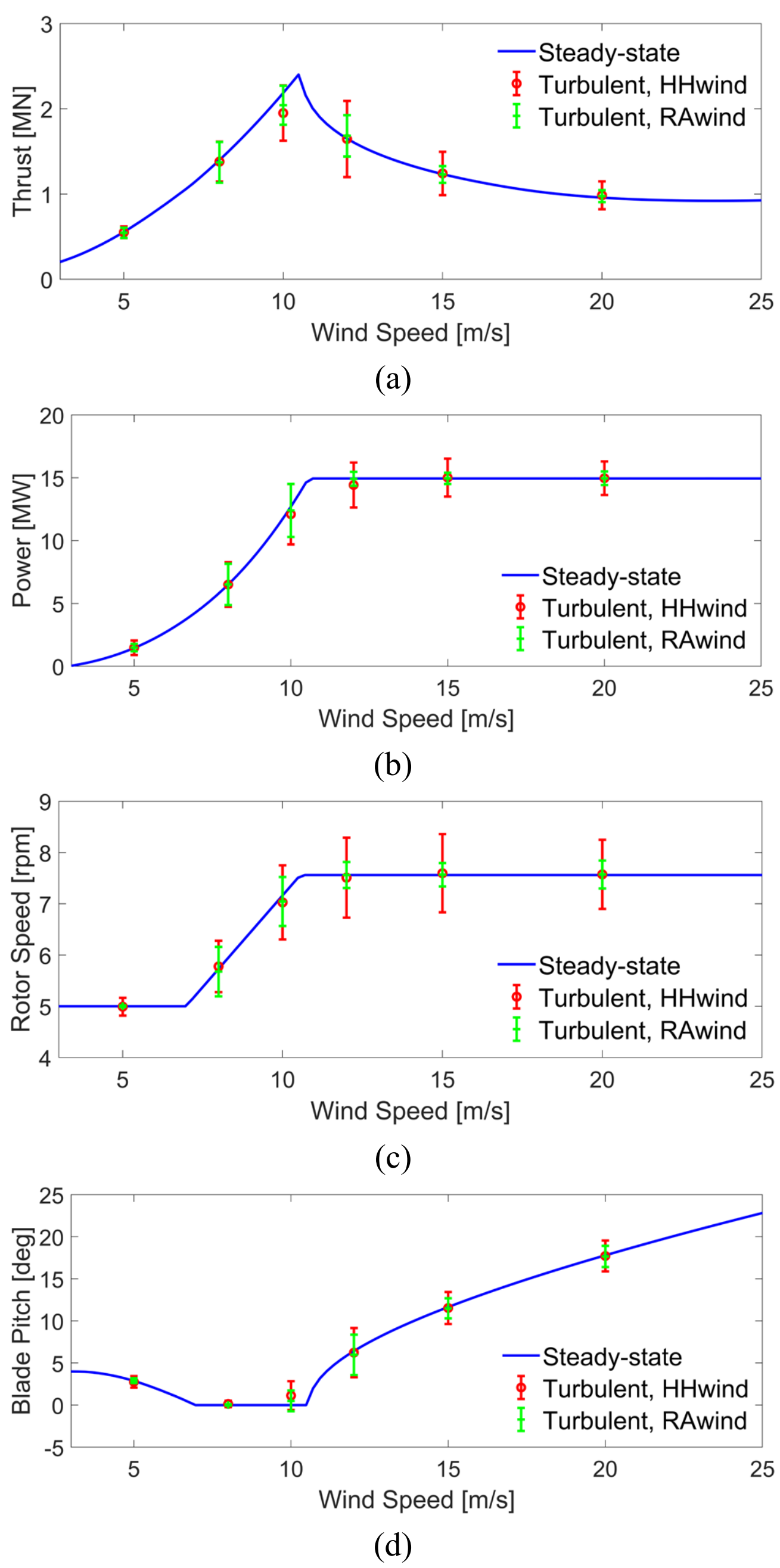

**Fig. 11.** Performance and operation for the bottom-fixed IEA 15 MW wind turbine under turbulent shear wind. (a) Wind thrust; (b) Generator power; (c) Rotor speed; (d) Blade pitch.

### *4.2. Floating wind turbine*

As mentioned in Section 2.3, we propose a modified control strategy for FOWTs. This considers a $k_\beta$ term with gain scheduling that adapts to the relative wind speed for blade pitch compensation and a $k_\tau$ term for generator torque compensation. An alternative strategy applies one constant $k_\beta$ across all wind speeds while omitting the $k_\tau$ term (hereafter as constant $k_\beta$ control), resembling ROSCO's implementation. In following simulations, we compare the performance of the modified control strategy with the baseline and constant $k_\beta$ control strategies, as detailed in Table 3. The constant $k_\beta$ value is heuristically set to 3.54 deg/(m/s).

Fig. 12 shows the variation of $k_\beta$ with wind speed according to Eq. (29), as denoted by the blue line. $k_\beta$ decreases as wind speed increases in the above-rated range, and it is fixed at 0 at below-rated wind speeds. To avoid the abrupt change of $k_\beta$ between operational regions, we implement a 10th-order polynomial a transition, shown by the green dash–dotted line in Fig. 12. The red dashed line indicates the constant $k_\beta$ of 3.54 deg/(m/s) employed in the constant $k_\beta$ control strategy.

**Table 3.** The control strategies considered in this study. (* Note the baseline control strategy here is different from ROSCO with the steady-state operating point being determined based on the relative wind speed instead of the inflow wind speed)

| Control strategy | $k_\beta$ | $k_\tau$ |
|---|---|---|
| Baseline* | 0.0 | 0.0 |
| Const. $k_\beta$ | 3.54 | 0.0 |
| Modified | Eq. (29) | Eq. (31) |

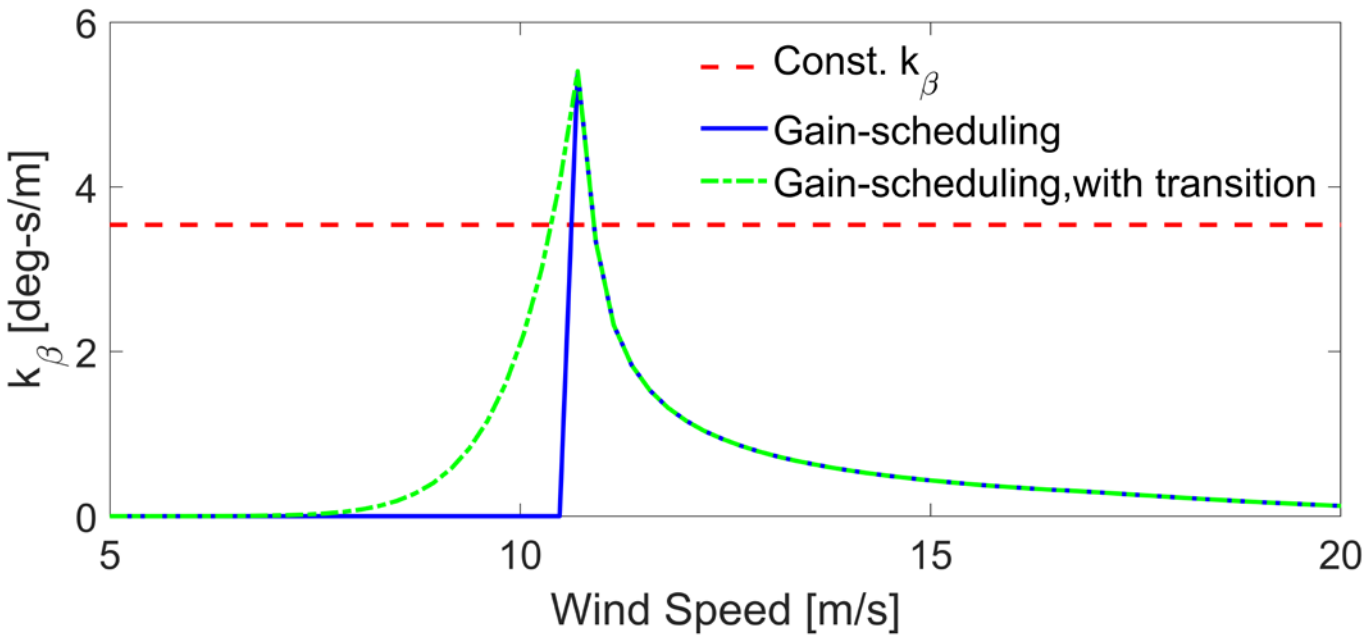

**Fig. 12.** The variation of $k_\beta$ with wind speed.

#### *4.2.1. Demonstration of negative damping*

In this subsection, we start with a simple case to demonstrate the negative damping phenomenon. The floating wind turbine is in still water without wave excitation and exposed to a steady uniform wind. Fig. 13 shows the results when the wind speed is 12 m/s with different control strategies. A ramp function is imposed on the thrust with a transition period of 100 s to avoid the transient effect on the platform motion at the beginning. An instability occurs with a remarkable pitch motion of the platform at natural frequency using the baseline control, which grows in a short duration and maintains large-amplitude oscillations afterwards. This is due to the negative damping effect that the derivative of thrust with respect to wind speed becomes negative at above-rated wind speeds, as verified by Fig. 13(b) and (d) that the wind thrust increases with the decrease of the relative wind velocity at the beginning. Nevertheless, when using the constant $k_\beta$ or modified control strategy, no instability issue occurs in platform pitch motion and the results can quickly reach the steady state of a constant thrust and a static offset in platform pitch. The corresponding blade pitch angle adjustment with time using these three control strategies is shown in Fig. 13(c). Note that the maximum blade pitch rate (in absolute value) is set to 8°/s here to highlight the negative damping phenomenon without incorporating much effect of blade pitch rate limit.

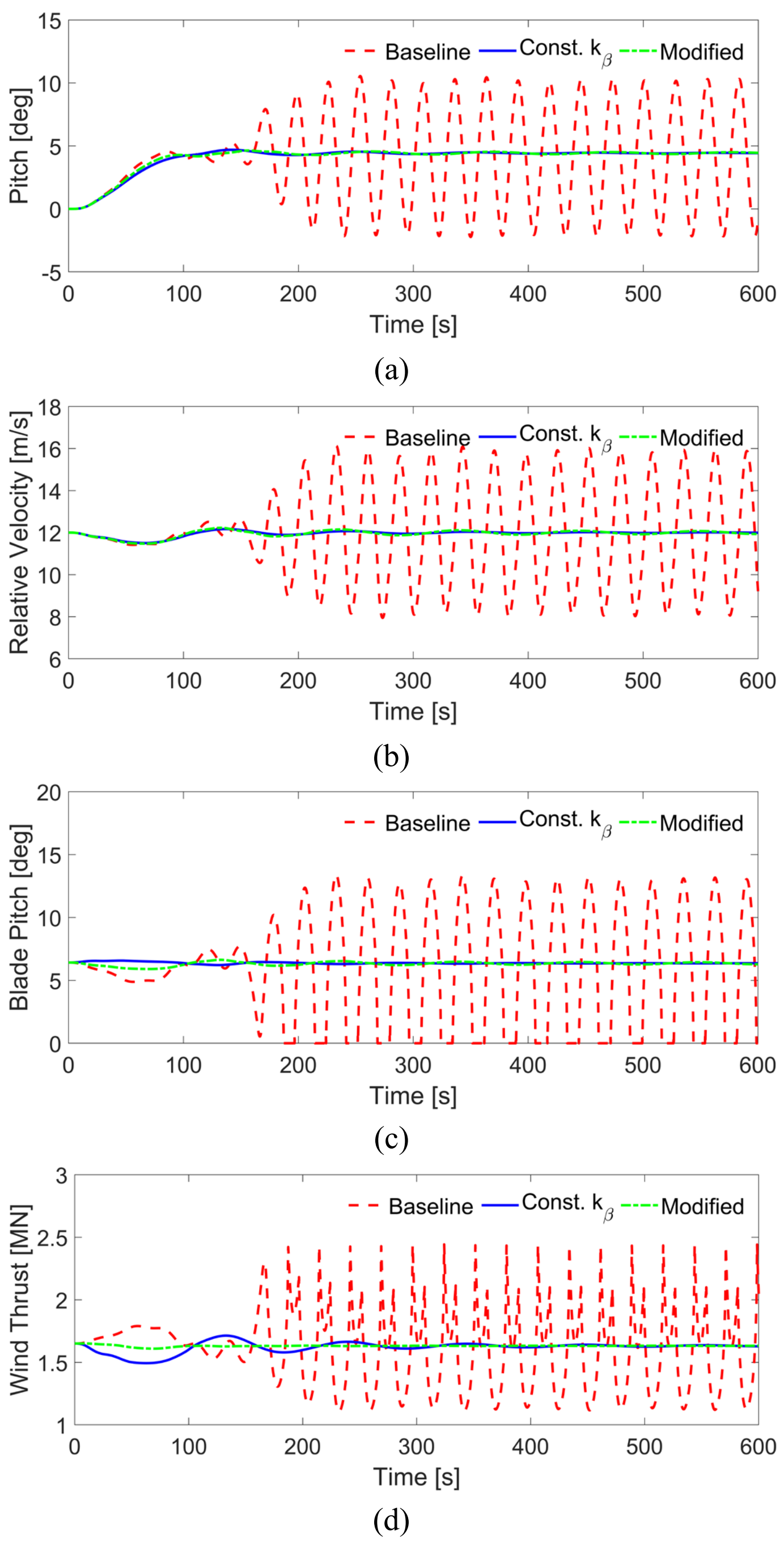

**Fig. 13.** Results for the FOWT under steady wind of 12 m/s with different control strategies. (a) Platform pitch; (b) Relative wind velocity at the hub; (c) Blade pitch; (d) Wind thrust.

Then the floating wind turbine in imposed motions is investigated. Here, a sinusoidal platform surge or pitch motion is imposed with an oscillation frequency $f$ and amplitude $A$. The dynamic variation of thrust is plotted as a function of hub velocity following Bayati et al. (2016). Fig. 14(a) and (b) shows the results for steady wind of 12 m/s with the imposed surge motion ($A$ = 1.0 m, $f$ = 0.1 Hz) and pitch motion ($A$ = 1.0 deg, $f$ = 0.036 Hz), respectively. For the imposed pitch motion, the turbine is rotated around the CoM of the system. In the figure, the elliptical curve slope quantifies aerodynamic damping (in-phase with the velocity), while the hysteresis area relates to aerodynamic added mass and stiffness. It is clear that at this wind speed, the aerodynamic damping is negative when using the baseline control without platform motion feedback. For the modified control strategy, the aerodynamic damping approaches zero as expected. For the constant $k_\beta$ control strategy, the aerodynamic damping is positive since a larger constant $k_\beta$ (see Fig. 12) is adopted so that $C_{\text{aero}} > 0$ in Eq. (28). It is interesting that more hysteretic behaviours are observed in imposed pitch motions. More test cases about the influence of control strategies on negative damping in different scenarios will be discussed in the next section.

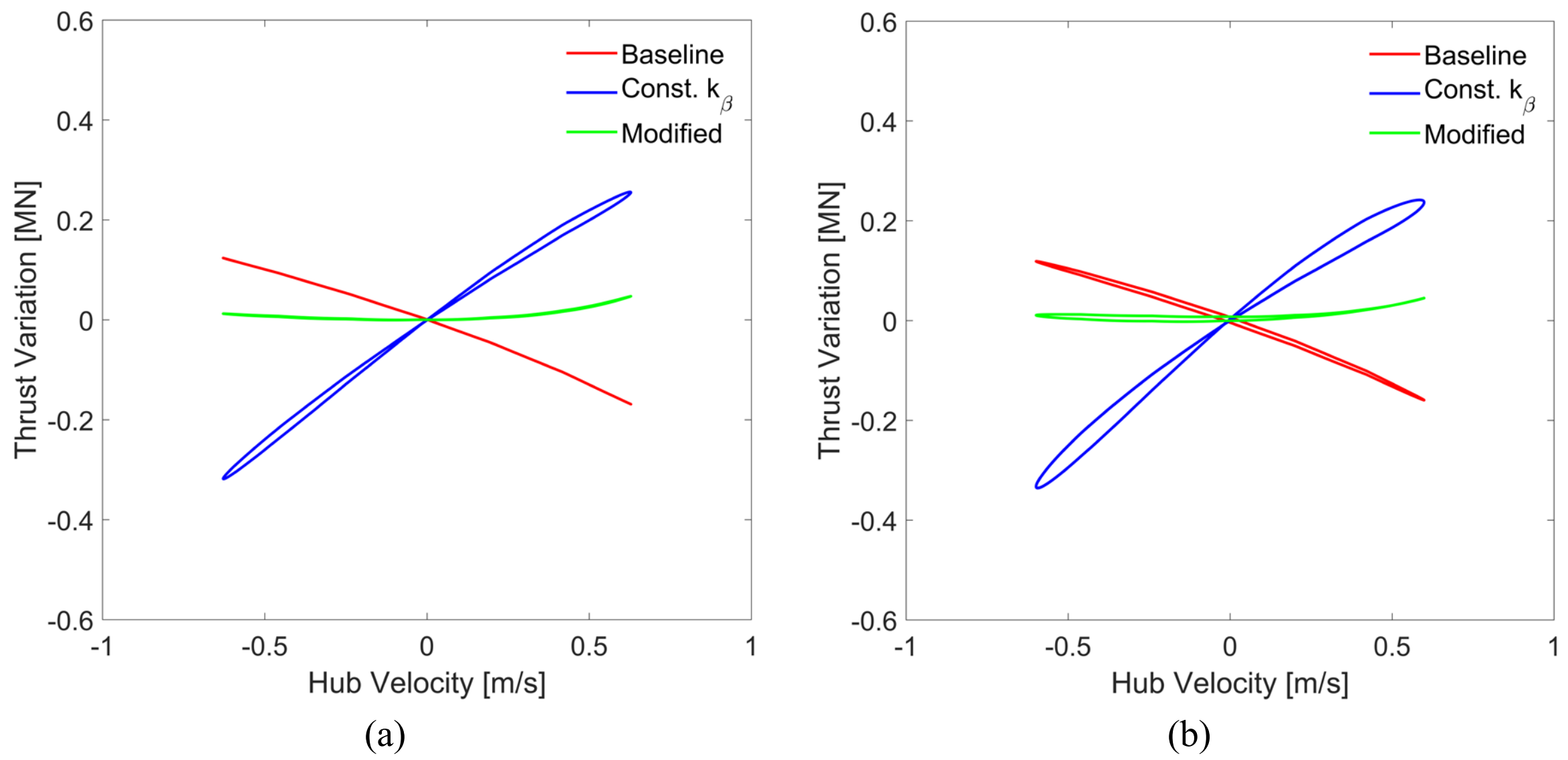

**Fig. 14.** Thrust variation with hub velocity for steady wind of 12 m/s using different control strategies. (a) Imposed surge motion ($A$ = 1.0 m, $f$ = 0.1 Hz); (b) Imposed pitch motion ($A$ = 1.0 deg, $f$ = 0.036 Hz).

### *4.2.2. Effects of control strategy*

#### *4.2.2.1 Steady wind and regular waves*

The steady uniform wind with different inflow wind speeds ($U_{\text{wind}}$) from 8 to 18 m/s is considered here. The regular waves of 5.0 m wave height and 10 s period are adopted for each wind speed. In this subsection, the maximum blade pitch rate (in absolute value) is set to 8°/s when using the baseline control to highlight the negative damping phenomenon without incorporating much effect of blade pitch rate limit.

Fig. 15 shows the results when $U_{\text{wind}}$ = 12 m/s with different control strategies. At this wind speed, the instability occurs with remarkable variations of the platform pitch, rotor speed, blade pitch and wind thrust with time due to the negative damping effect when using the baseline control, which will be further confirmed in Fig. 18(a). Nevertheless, when using the constant $k_\beta$ or modified control strategy, no instability issue occurs and the results have relatively small and stable oscillations around the mean value arising from the regular wave excitation. Furthermore, the modified control strategy produces smallest variations of rotor speed and wind thrust as well as a smallest maximum wind thrust. Reducing the variation of rotor speed is of great significance to the smooth operation of the rotor. The huge differences in platform pitch motion between the baseline control and the other two control strategies are from the significant response at natural frequency of around 0.036 Hz, as illustrated by the amplitude spectra shown in Fig. 17(a). For the wave-frequency response at 0.1 Hz, the amplitude is the smallest when using the constant $k_\beta$ control strategy, as it has the largest $k_\beta$ among three control strategies around $U_{\text{wind}}$ = 12 m/s (see Table 3 and Fig. 12), which increases the system damping. Note the mean offset is subtracted from the results in spectral analysis so there is no component at zero frequency.

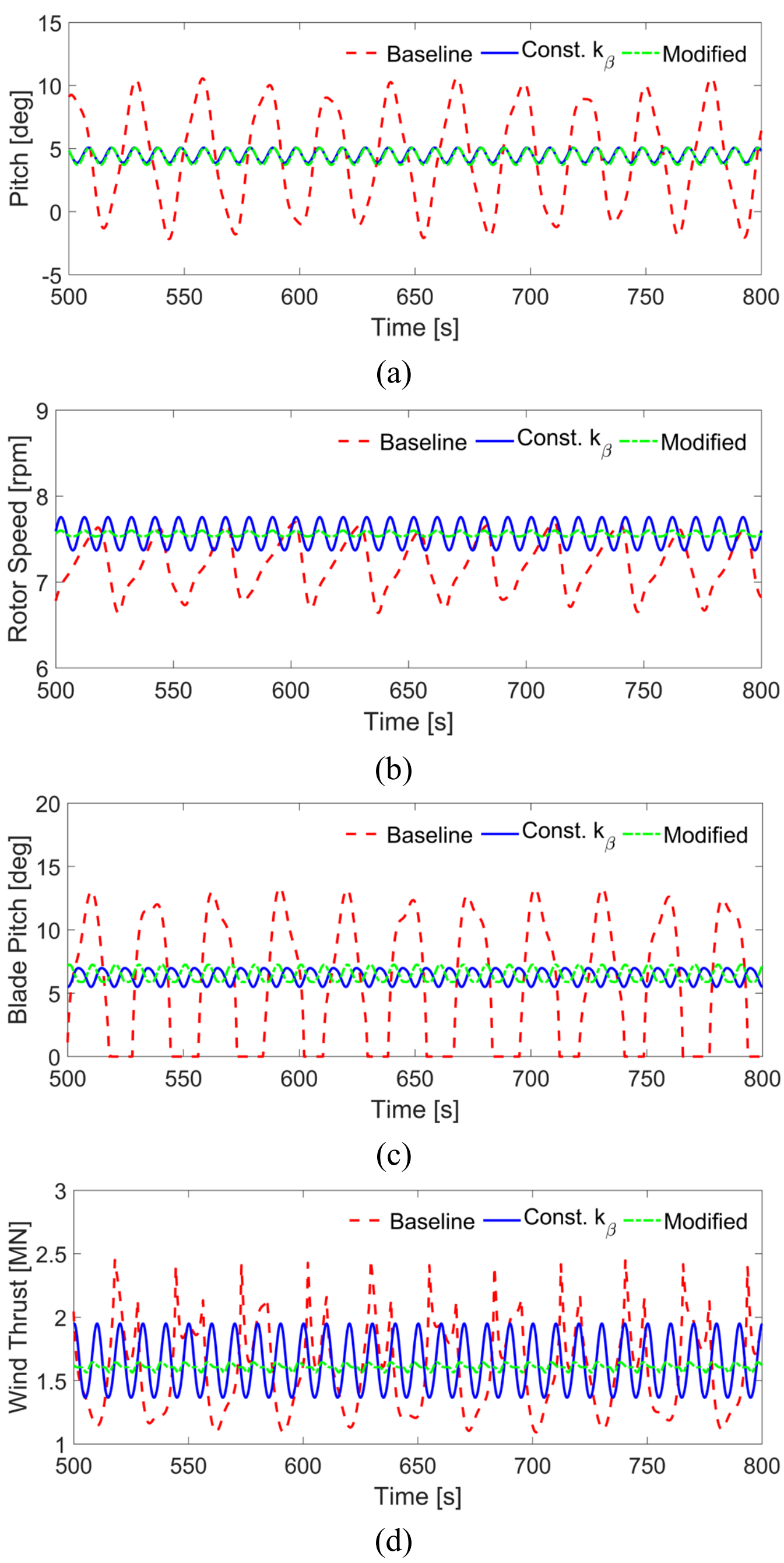

**Fig. 15.** Results for the FOWT under steady wind of 12 m/s and regular waves with different control strategies. (a) Platform pitch; (b) Rotor speed; (c) Blade pitch; (d) Wind thrust.

Fig. 16 shows the results when $U_{wind}$ = 16 m/s, and no obvious instability occurs at this wind speed for any of the three control strategies. This is because even using the baseline control, the aerodynamic damping is negative but its magnitude is small around $U_{wind}$ = 16 m/s, as will be verified in Fig. 18(b). The overall damping of the FOWT system remains positive plus a certain level of hydrodynamic damping, and thus the instability of platform pitch motion is not triggered and does not grow. Similar results are obtained by using the baseline and modified control strategies, as the $k_\beta$ with gain scheduling is quite small around $U_{wind}$ = 16 m/s. The constant $k_\beta$ strategy produces a smaller platform pitch amplitude at wave frequency due to an obviously larger $k_\beta$ (see Fig. 12), which is further illustrated by Fig. 17(b). However, it leads to much larger variations of rotor speed, blade pitch and wind thrust with time. This is not conducive to the smooth operation of the rotor, and may cause large rotor overspeed.

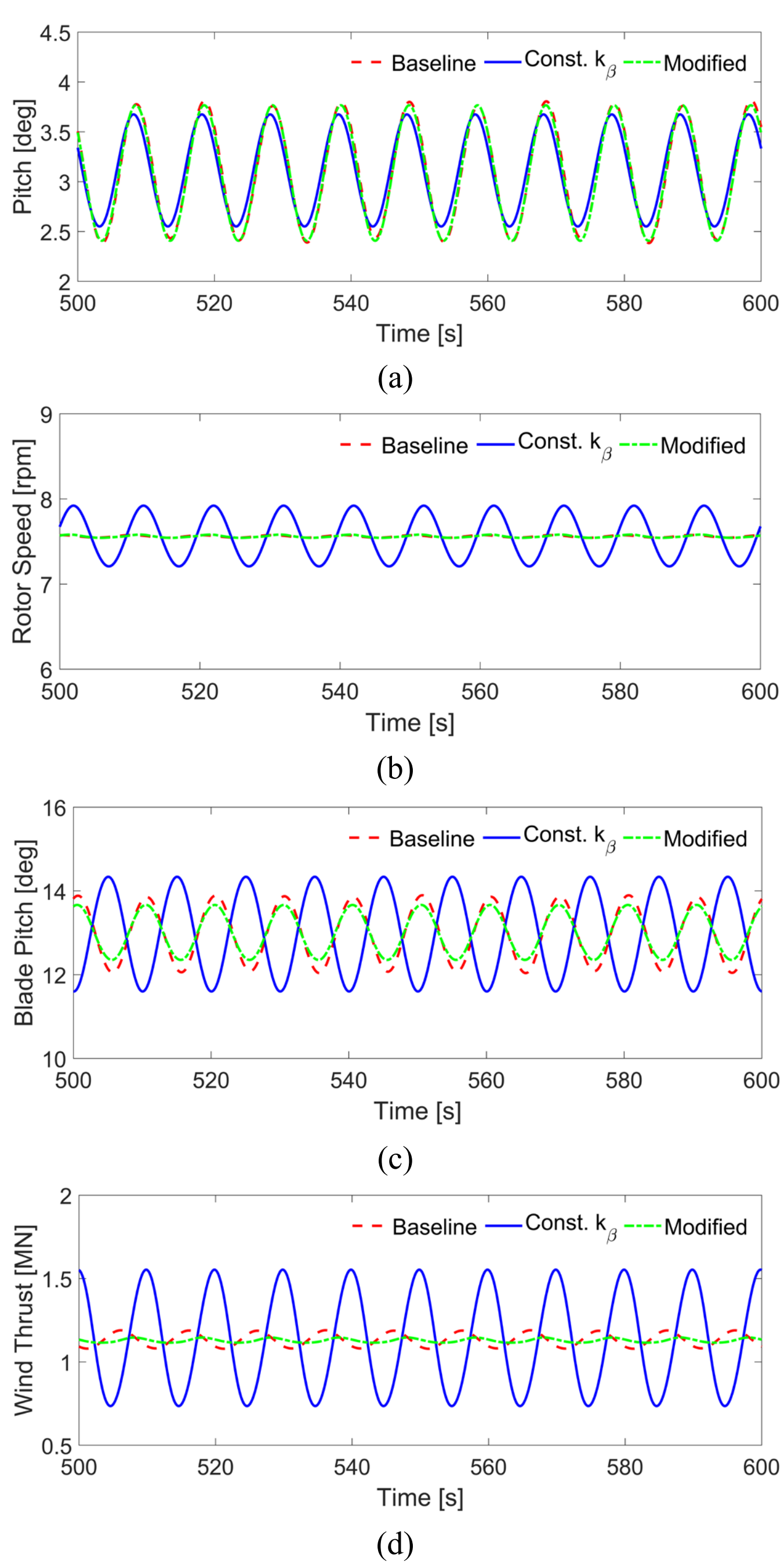

**Fig. 16.** Results for the FOWT under steady wind of 16 m/s and regular waves with different control strategies. (a) Platform pitch; (b) Rotor speed; (c) Blade pitch; (d) Wind thrust.

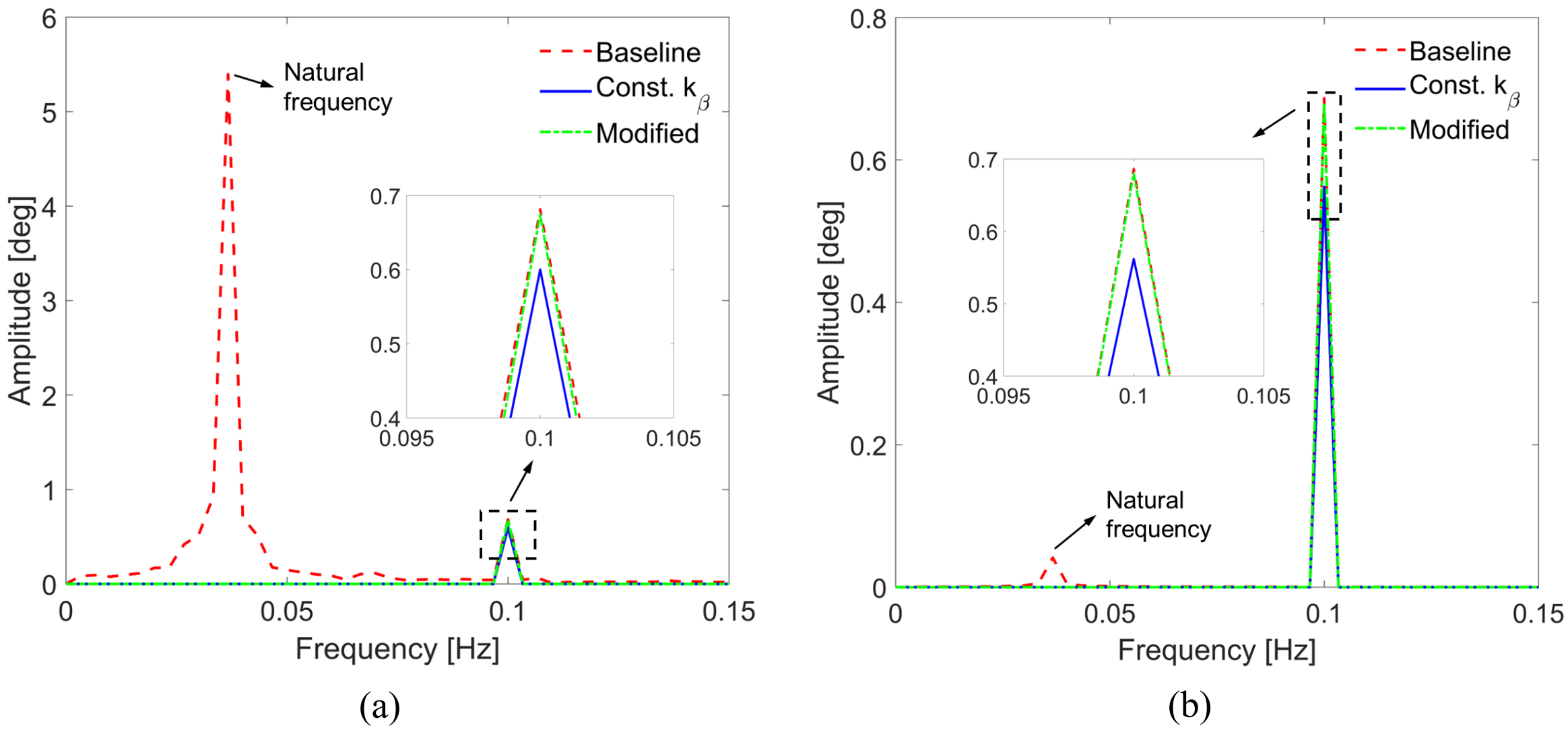

**Fig. 17.** Amplitude spectra of platform pitch with different control strategies. (a) $U_{wind}$ = 12 m/s; (b) $U_{wind}$ = 16 m/s.

To further demonstrate the aerodynamic damping with different control strategies under different wind speeds, Fig. 18(a) and (b) shows the dynamic variation of thrust versus hub velocity with $U_{wind}$ = 12 and 16 m/s respectively. For $U_{wind}$ = 12 m/s, it is seen that if using the baseline control the slope (aerodynamic damping) is negative when the relative wind velocity is above-rated and reaches the minimum near-rated, while it becomes positive below rated wind speed. The aerodynamic damping approaches zero using the modified control as expected, and it keeps positive for using the constant $k_\beta$ control strategy since a larger constant $k_\beta$ is adopted so that $C_{\text{aero}} > 0$ in Eq. (28). For $U_{wind}$ = 16 m/s, the negative aerodynamic damping is quite small even using the baseline control, and the trends of aerodynamic damping for the modified control and the constant $k_\beta$ control are similar to those when $U_{wind}$ = 12 m/s.

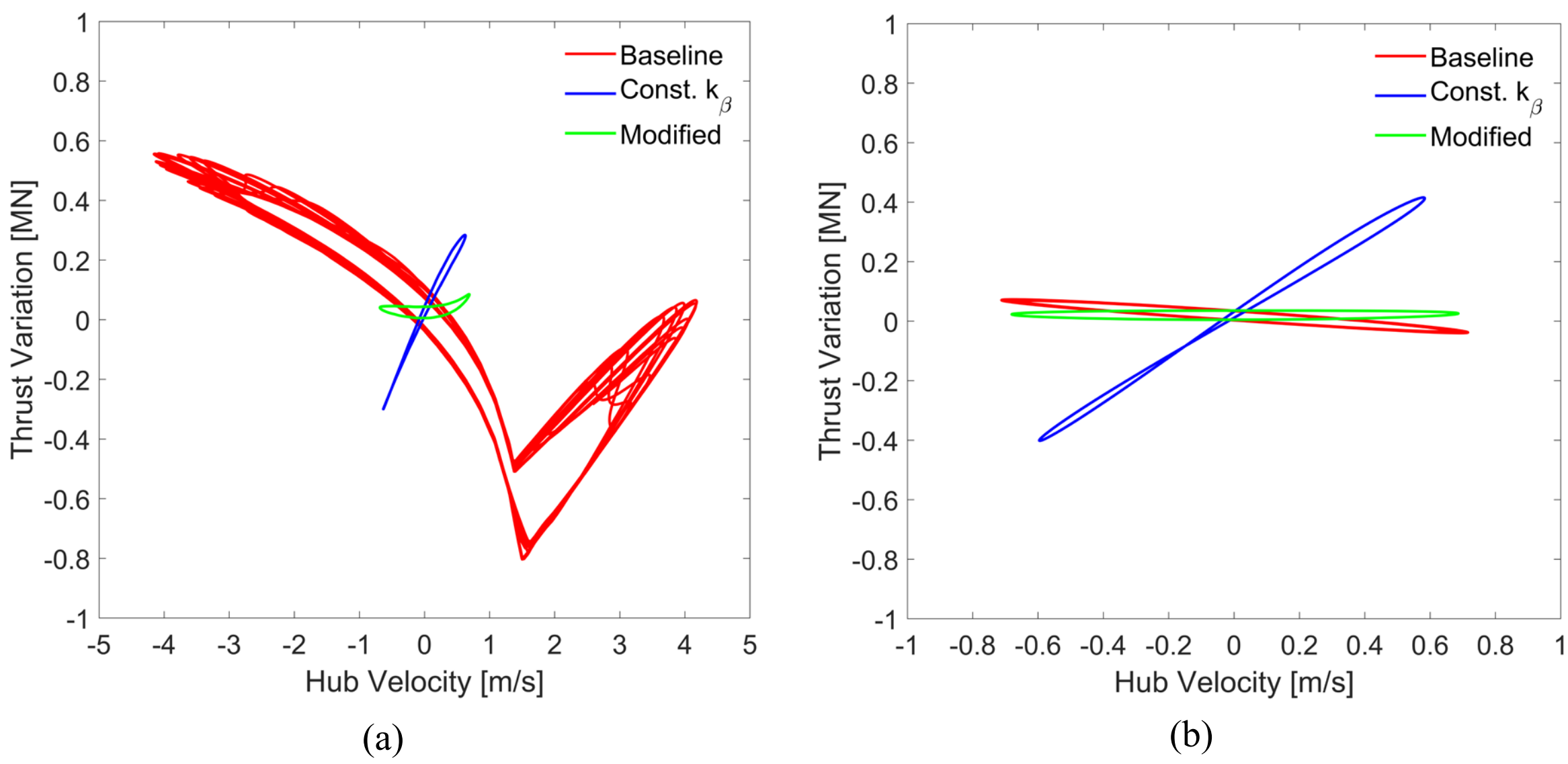

**Fig. 18.** Thrust variation with hub velocity using different control strategies. (a) $U_{wind}$ = 12 m/s; (b) $U_{wind}$ = 16 m/s.

Fig. 19 shows the platform pitch statistics at different wind speeds. Note the statistics are obtained from 300 s (30 wave periods) time series after the results become stable. It is notable that the constant $k_\beta$ and modified control strategies significantly reduce the standard deviation (std) and maximum of platform pitch when $U_{wind}$ is between 11 m/s and 14 m/s, as these two control strategies increase the system damping and eliminate the negative damping effect that occurs when using the baseline control at these wind speeds. Beyond the above wind speed range, the results of std and maximum are slightly smaller when using the constant $k_\beta$ control but the difference is small using different control strategies. The platform pitch average is similar except for that at wind speed of 11 m/s where the average is the smallest when using the baseline control. This is because the average of wind thrust is the smallest using the baseline control at this wind speed though not shown here. Fig. 20 presents the maximum wind thrust at different wind speeds. The modified control yields the smallest maximum wind thrust at all above-rated wind speeds, while the values are quite close for below-rated wind speeds. Compared to the baseline control, the constant $k_\beta$ control reduces the maximum thrust at wind speeds when negative damping occurs, while it increases the maximum thrust at larger wind speeds.

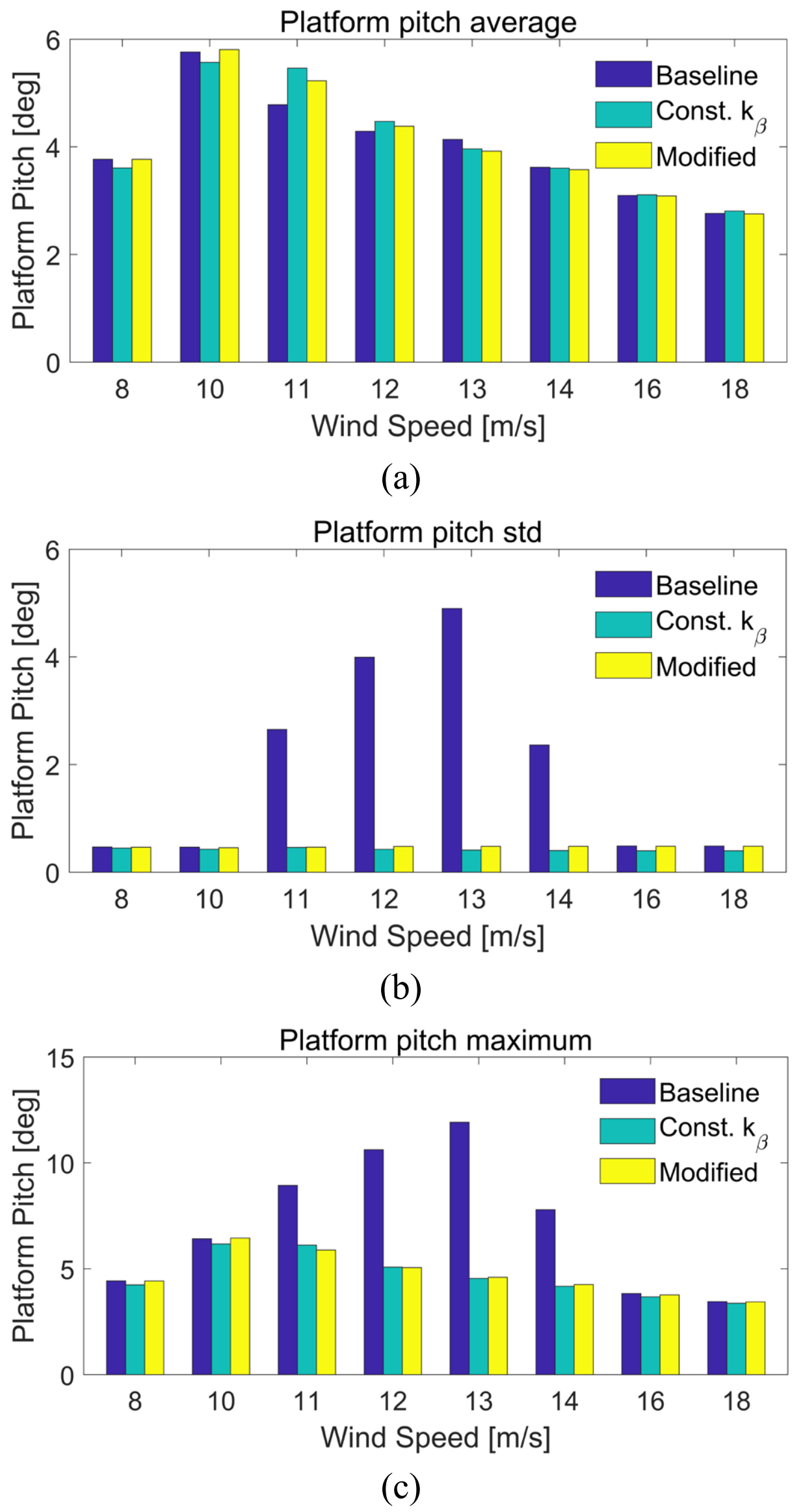

**Fig. 19.** Platform pitch statistics for the FOWT under steady wind and regular waves with different control strategies. (a) Platform pitch average; (b) Platform pitch std; (c) Platform pitch maximum.

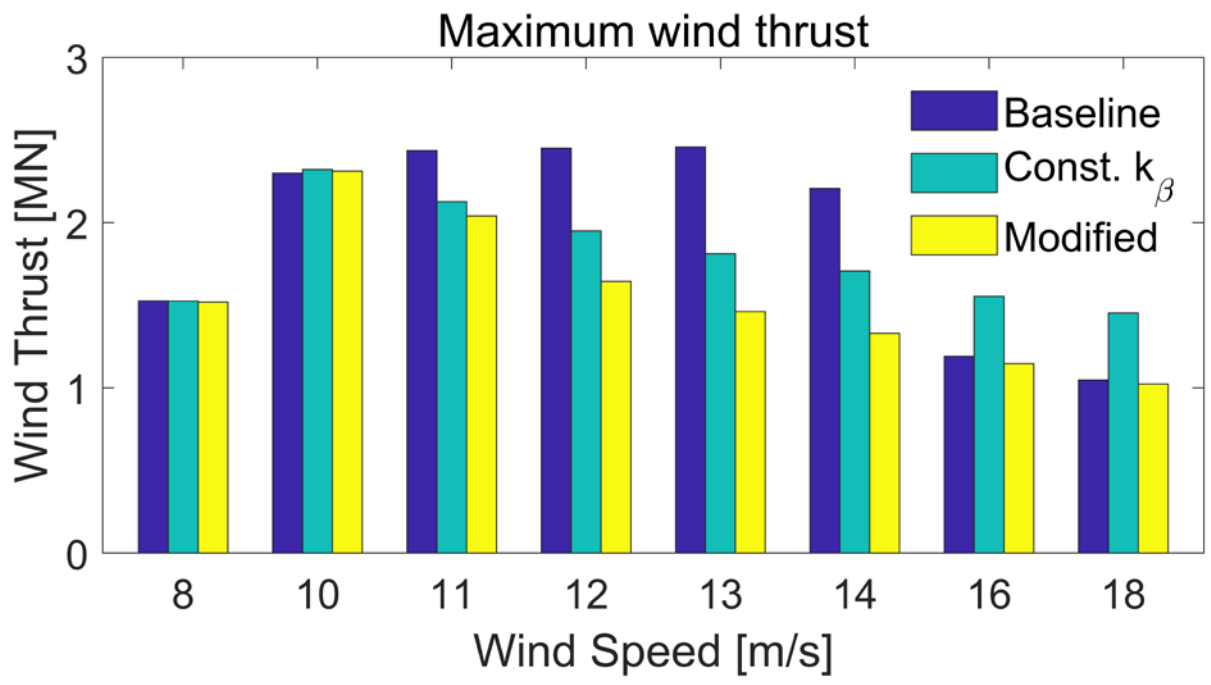

**Fig. 20.** Maximum wind thrust for the FOWT under steady wind and regular waves with different control strategies.

Fig. 21 shows effects of the control strategy on the rotor performance at different wind speeds. It is worth noting that the modified control yields the smallest rotor speed std at almost all wind speeds except for $U_{wind}$ equal to and larger than 16 m/s where the rotor speed std is similar with that using the baseline control. The large rotor speed std using the baseline control when $U_{wind}$ is between 11 and 14 m/s is mainly due to the significant variation of relative wind velocity with time caused by the negative damping effect. Though the constant $k_\beta$ control eliminates the negative damping effect and reduces the platform pitch, it produces a much larger rotor speed std compared to the modified control. It is interesting that the rotor speed std increases with increasing wind speed in above-rated range when the constant $k_\beta$ control is used, which is unfavourable to the rotor's smooth operation and enlarges the maximum rotor speed at large wind speeds (see Fig. 21(c)). The averages of rotor speed and generator power are almost the same when using different control strategies except that the averages reduce when the negative damping occurs, as shown in Fig. 21(a) and (d). Overall, the modified control strategy shows the superiority in reducing both the platform motion and rotor speed variation for all wind speeds considered.

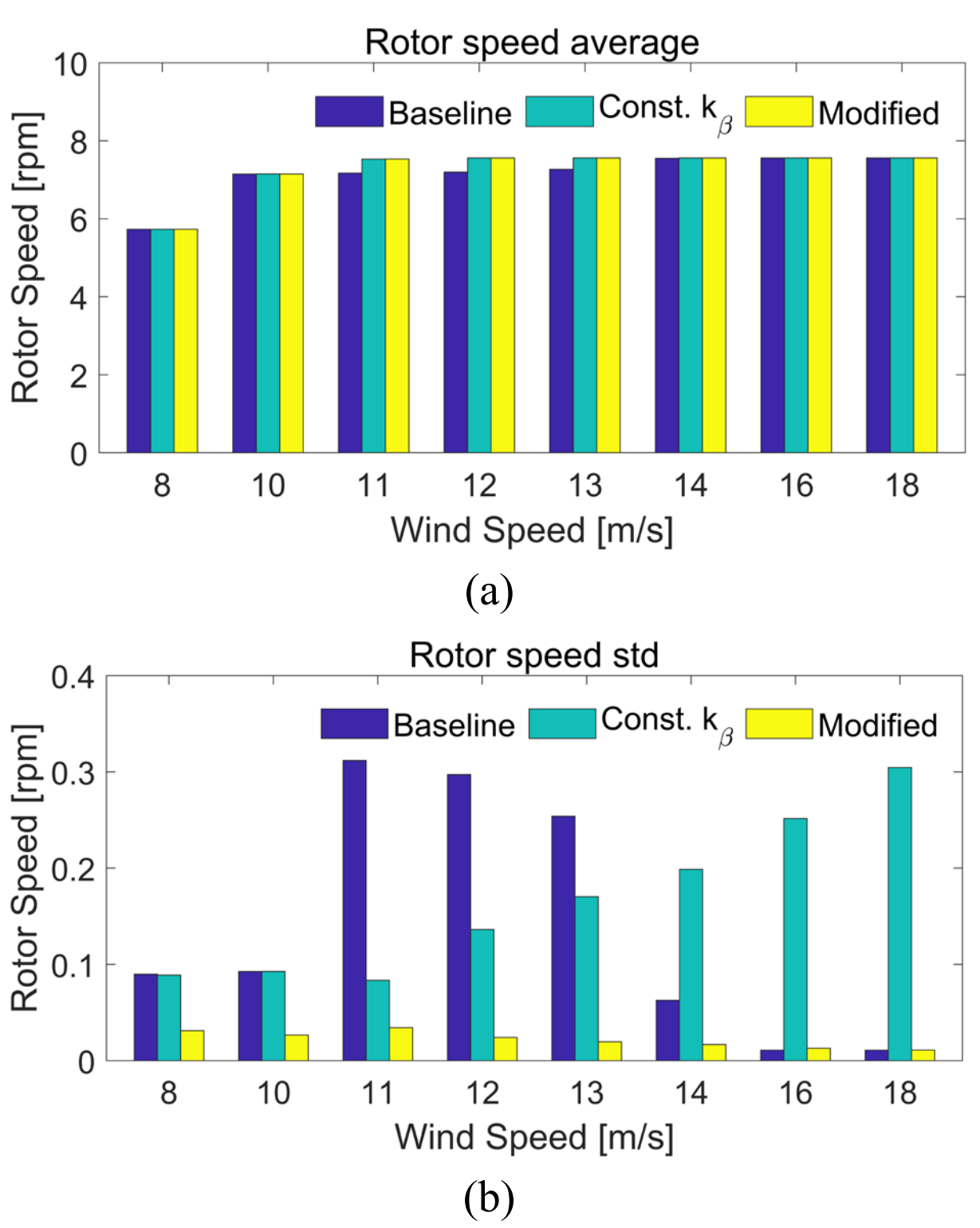

(a)

(b)

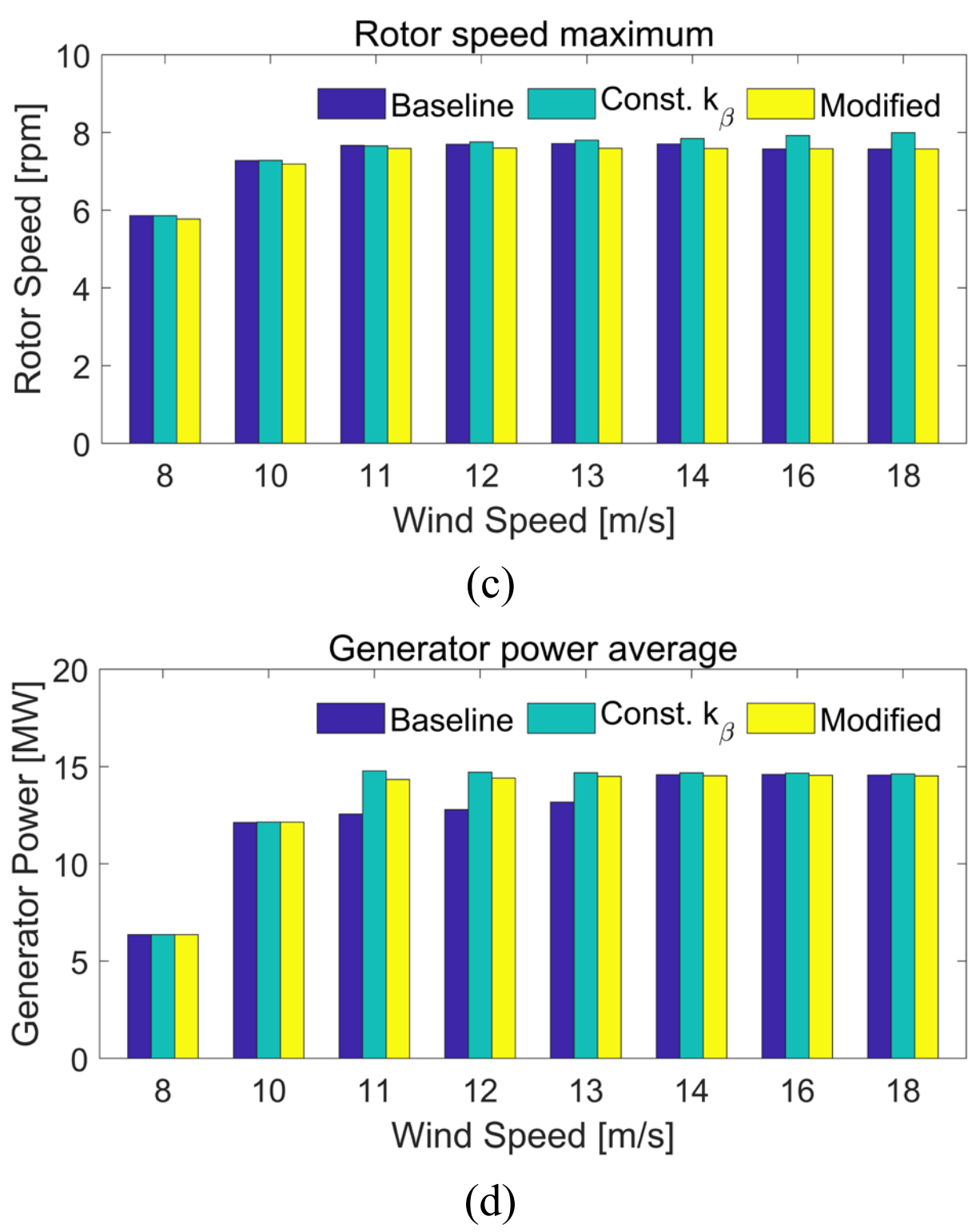

**Fig. 21.** Rotor performance for the FOWT under steady wind and regular waves with different control strategies. (a) Rotor speed average; (b) Rotor speed std; (c) Rotor speed maximum; (d) Generator power average.

*4.2.2.2 Turbulent wind and irregular waves*

In this subsection, a subset of IEC design load cases (DLCs) is chosen to incorporate representative normal design conditions. The metocean data for DLC 1.6 (severe sea states) from Allen et al. (2020) are adopted, as listed in Table 4. Note that all conditions consider an aligned wind and wave heading of 0˚. The normal turbulence model (NTM) is used with a standard IEC turbulence category of "B", and the JONSWAP spectrum is adopted for describing the irregular waves with spectral peakedness of 2.75. For each case, it runs for a total of 3600 s (one-hour sea state) with a random seed. It is shown that the present coupled aero-hydro-mooring-servo model is highly efficient. The computational time is around 10 min on a laptop for one-hour simulations. The simulation results are presented as average, standard deviation (std) and maximum values.

**Table 4.** Mean hub-height wind speed and wave parameters for the cases in DLC 1.6 (Allen et al., 2020). ($\bar{u}_{\mathrm{hub}}$: mean hub-height wind speed, Hs: significant wave height, $T_p$: peak period)

| Case ID | 1 | 2 | 3 | 4 | 5 | 6 | 7 |
|---|---|---|---|---|---|---|---|
| $\bar{u}_{\mathrm{hub}}$ (m/s) | 8 | 10 | 12 | 14 | 16 | 18 | 20 |
| $H_s$ (m) | 8.0 | 8.1 | 8.5 | 8.5 | 9.8 | 9.8 | 9.8 |
| $T_p$ (s) | 12.7 | 12.8 | 13.1 | 13.1 | 14.1 | 14.1 | 14.1 |

Fig. 22 shows the platform pitch statistics for different cases. Note that the horizontal axis of the bar chart is represented by $\bar{u}_{\mathrm{hub}}$ as it is unique and related to wave parameters for each case. Similar to the trends shown in Section 4.2.2.1, the constant $k_\beta$ and modified control strategies clearly reduce the platform pitch std and maximum when $\bar{u}_{\mathrm{hub}}$ is between 10 and 16 m/s compared to the baseline control, while the average values obtained by these control strategies are close to each other. In these cases, the negative damping effect results in large pitch response around the natural frequency when using the baseline control, which is confirmed by the power spectral density (PSD) shown in Fig. 23 with $\bar{u}_{\text{hub}}$ = 12 and 14 m/s for example. The constant $k_\beta$ and modified control strategies still works in mitigating the negative damping effect under these conditions.

Fig. 24 shows the rotor speed and generator power statistics for different cases. For the average values, the results are close to each other with different control strategies. The modified control has the advantage of reducing the std and maximum values of rotor speed at above-rated wind speed cases compared to the constant $k_\beta$ control. Note large rotor or generator overspeed spikes (~20% above rated) may trigger wind turbine shutdowns. Fig. 25 shows the variation of overspeed ratio with mean hub-height wind speed using different control strategies. It is seen that the overspeed ratio is below the safety threshold of 20% for all wind speeds considered in Table 4 when using the modified control, whereas it exceeds the threshold at large wind speeds ($\bar{u}_{\text{hub}}$ > 16 m/s) when using the constant $k_\beta$ control.

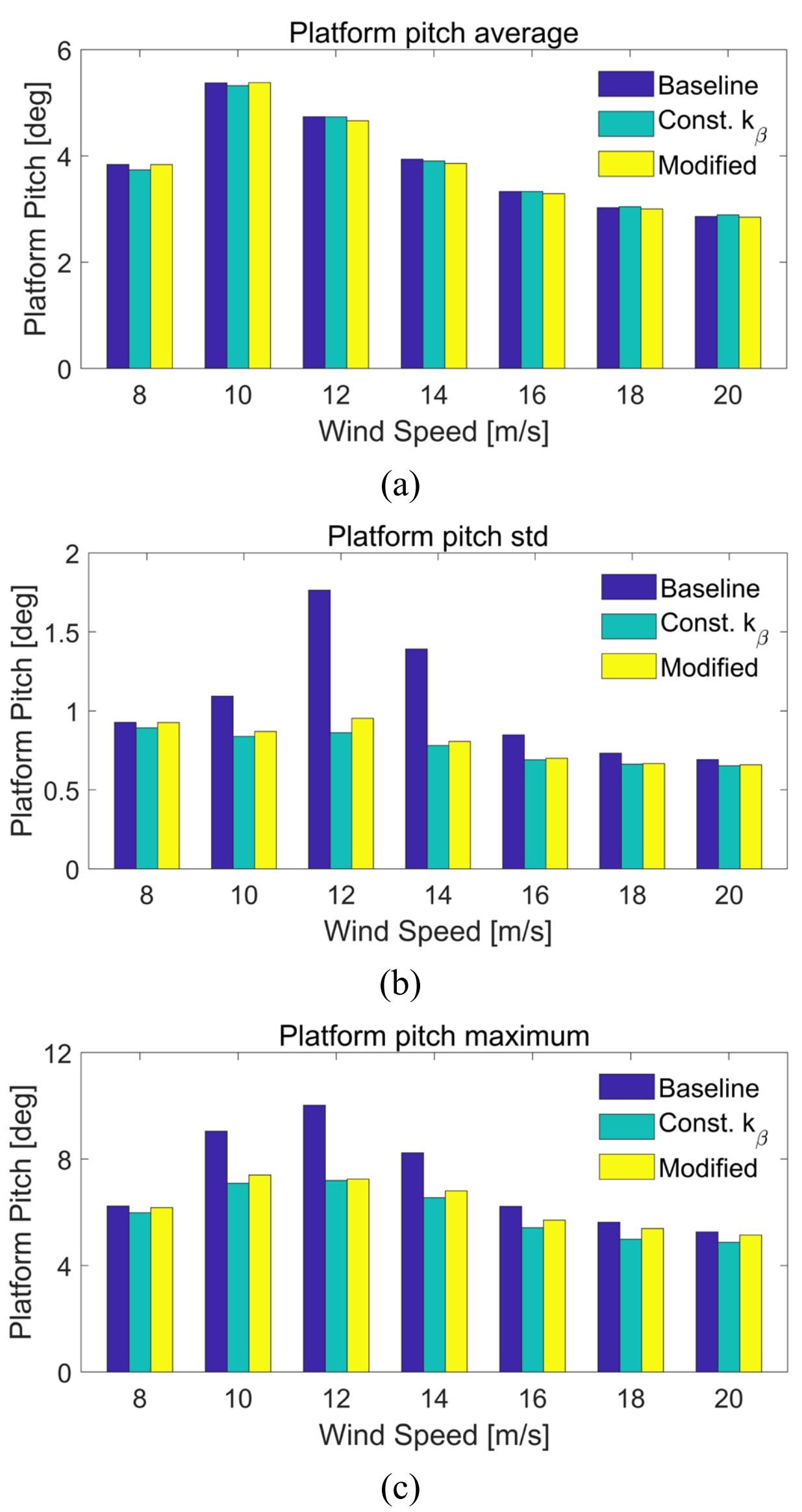

**Fig. 22.** Platform pitch statistics for the FOWT under turbulent wind and irregular waves with different control strategies. (a) Platform pitch average; (b) Platform pitch std; (c) Platform pitch maximum.

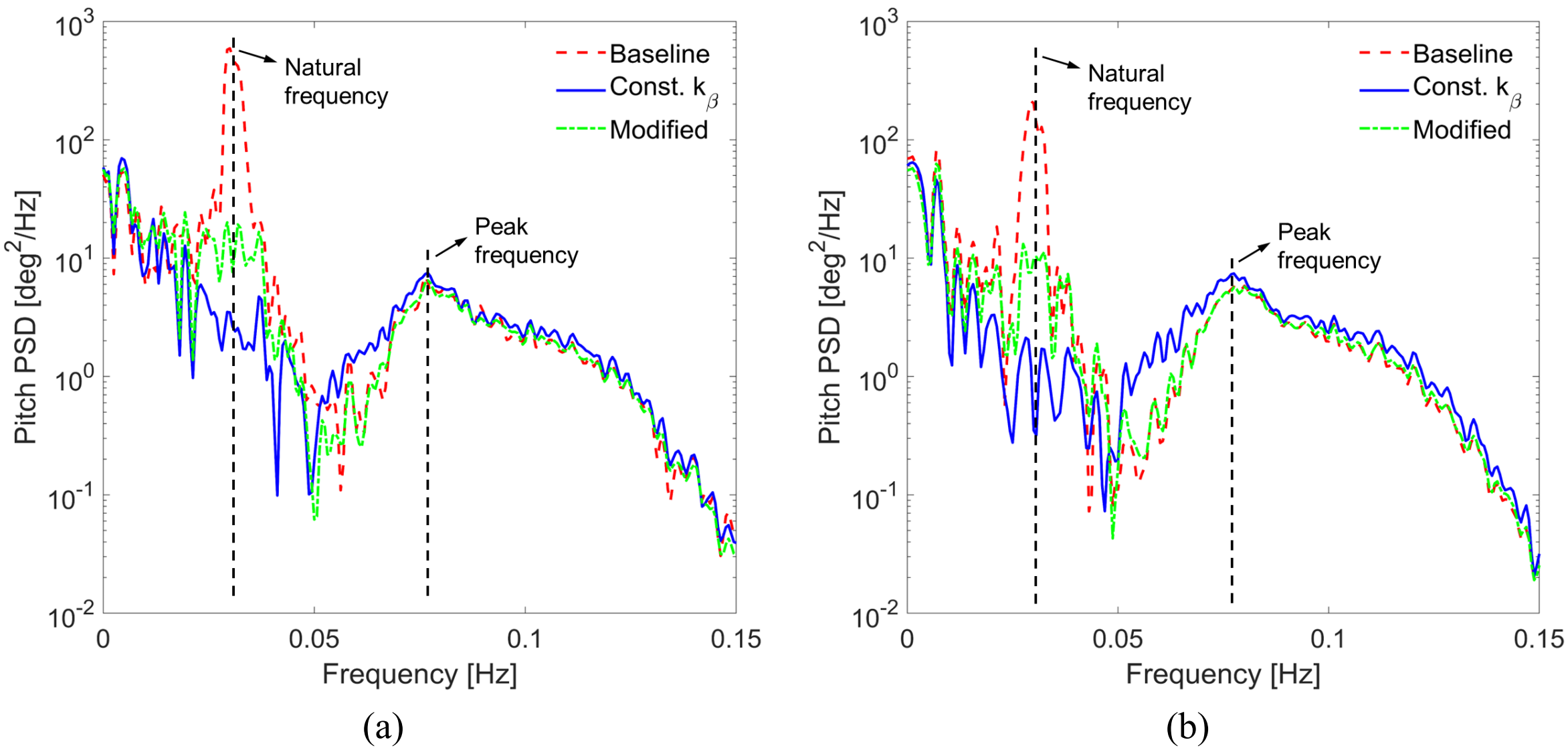

**Fig. 23.** Platform pitch PSD for the FOWT under turbulent wind and irregular waves with different control strategies. (a) $\bar{u}_{\text{hub}}$ = 12 m/s; (b) $\bar{u}_{\text{hub}}$ = 14 m/s.

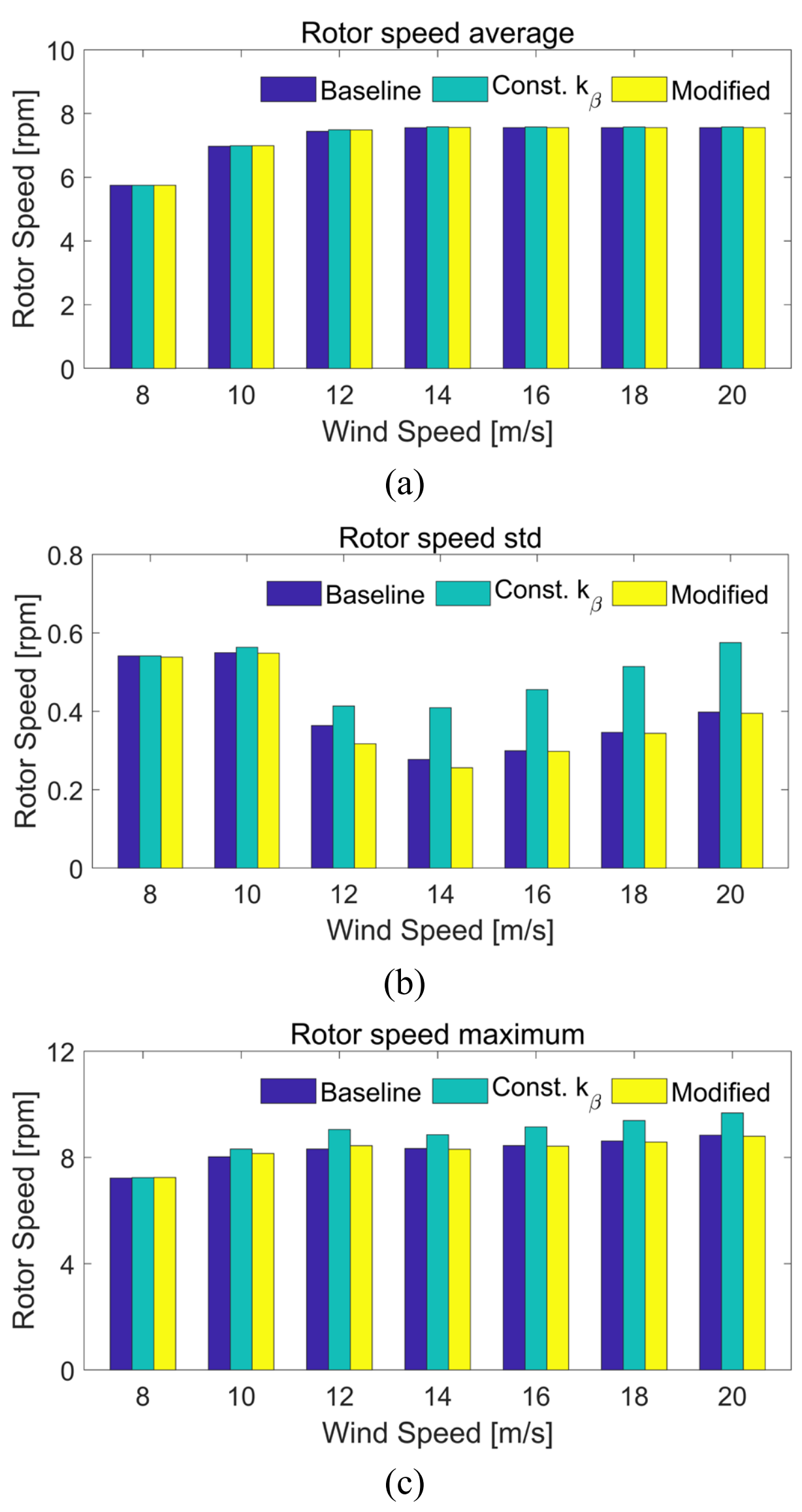

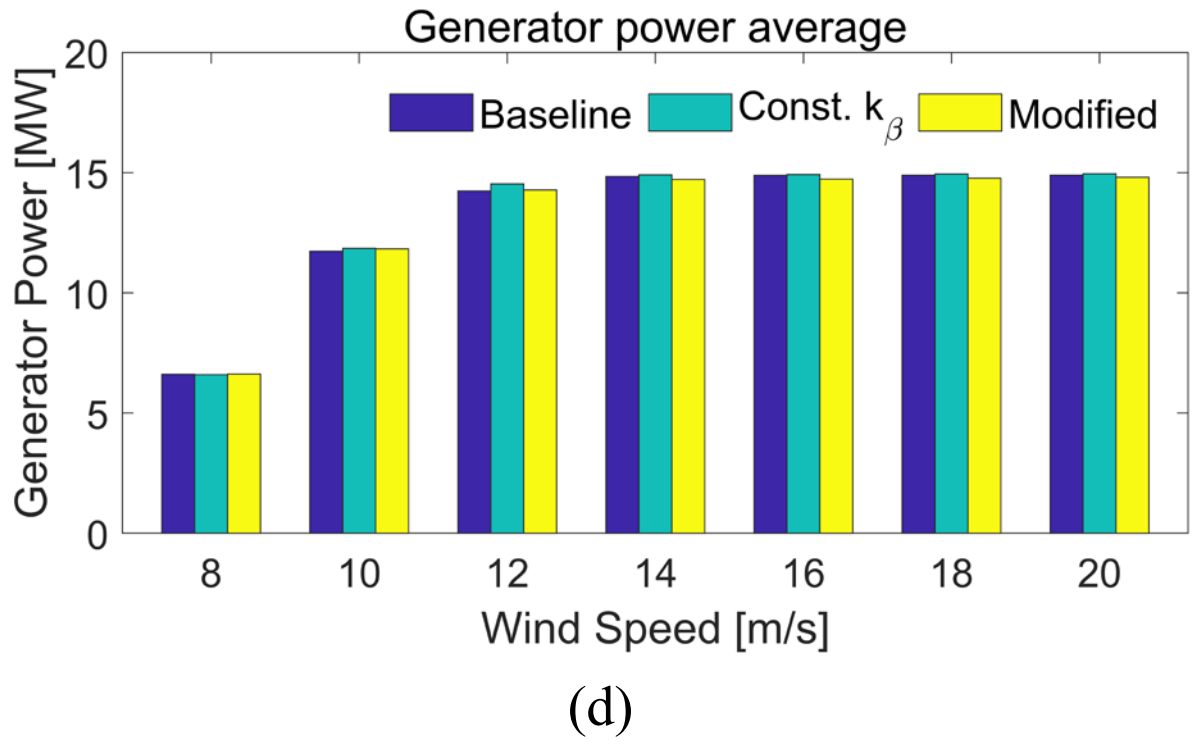

(d)

**Fig. 24.** Rotor performance for the FOWT under turbulent wind and irregular waves with different control strategies. (a) Rotor speed average; (b) Rotor speed std; (c) Rotor speed maximum; (d) Generator power average.

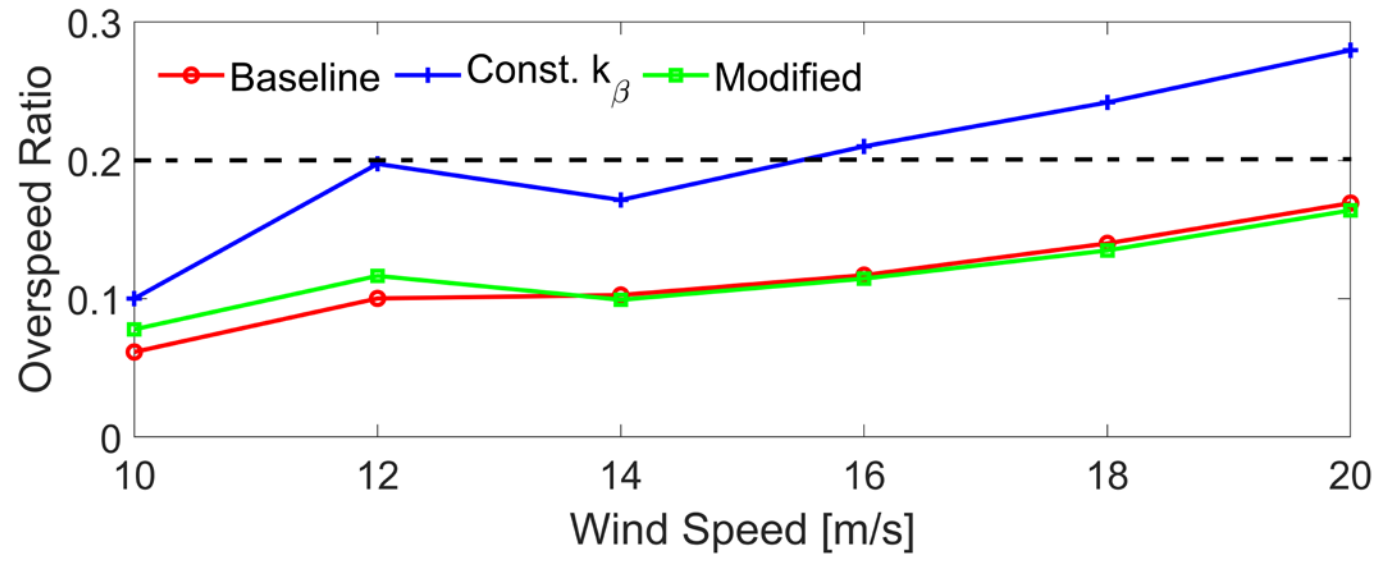

**Fig. 25.** Variation of overspeed ratio with mean hub-height wind speed using different control strategies. The black dashed line denotes the safety threshold of 20%.

### *4.2.3. Effects of turbulent wind*

In this subsection, the test cases listed in Table 4 are re-run under steady uniform wind conditions for comparison with turbulent wind results in Section 4.2.2.2 to independently assess the effects of turbulent wind on the platform motion and rotor performance. The incoming wind speed and wave parameters are the same as those in Section 4.2.2.2 for each case, and the modified control strategy is adopted for rotor control.

It is found that the main difference of the platform motion between turbulent and steady wind results lies in the low-frequency domain. Fig. 26 shows the platform pitch PSD for the wind speed of 10, 14 and 18 m/s. It is seen that turbulent wind results yield significantly larger low-frequency responses compared to steady wind results, while the difference of the responses within the wave-frequency range is relatively slight. Similar trends can also be observed in the surge motion of the platform though not shown here for brevity.

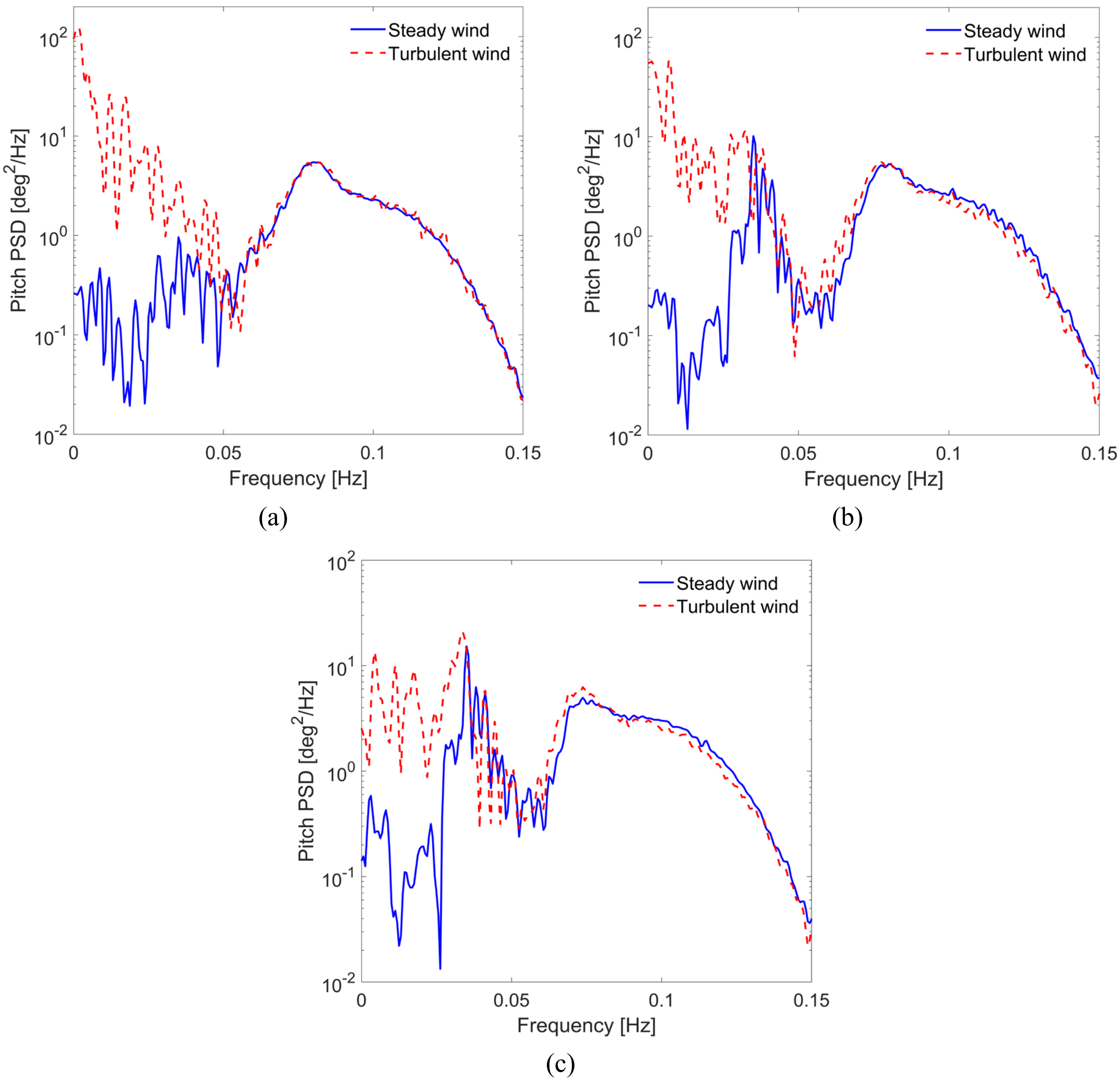

**Fig. 26.** Platform pitch PSD for the FOWT under irregular waves with different wind input. (a) $\bar{u}_{\text{hub}}$ = 10 m/s; (b) $\bar{u}_{\text{hub}}$ = 14 m/s; (b) $\bar{u}_{\text{hub}}$ = 18 m/s.

Fig. 27 presents the platform pitch statistics for these cases with different wind input. Turbulent wind results have larger platform pitch std and maximum than steady wind results as expected, and they have slightly larger average values except for the case with the wind speed of 10 m/s. It is interesting that the difference of platform pitch std between turbulent and steady wind results basically reduces with the increase of wind speed. This is generally consistent with the difference of wind thrust std between these two wind inputs, as illustrated in Fig. 28. The reason is that turbulent wind gives a wide range of relative wind speed during the simulation and thus leads to a larger thrust variation, and the thrust variation tends to flatten as wind speed increases.

Fig. 29 shows turbulent wind has little effect on the average values of rotor speed and generator power. Nevertheless, turbulent wind leads to a far greater rotor speed std (about from 6.8 to 16.2 times of steady wind results) and increases the maximum rotor speed by 7.9% to 23.7% for the wind speeds considered.

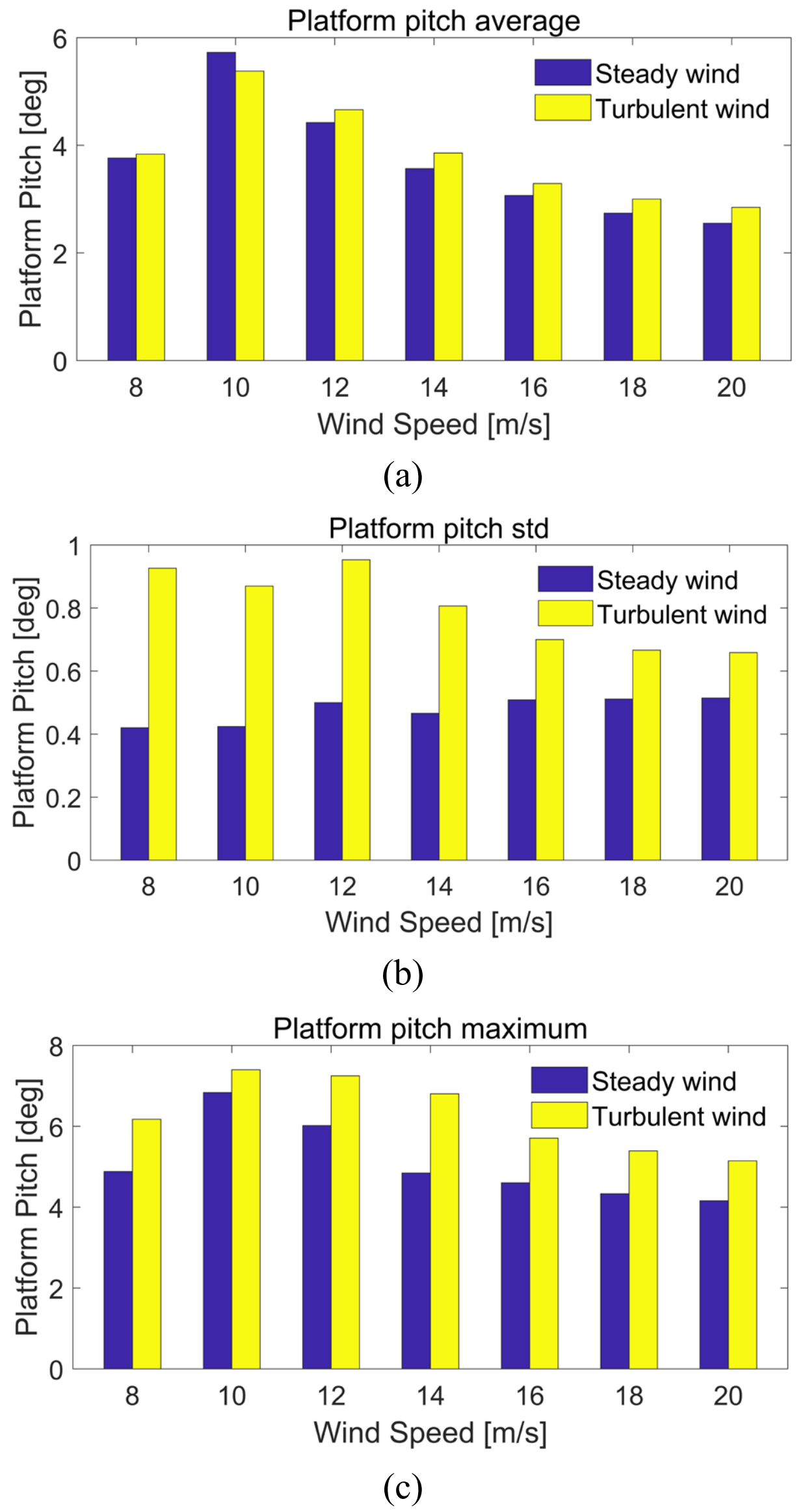

**Fig. 27.** Platform pitch statistics for the FOWT under irregular waves with different wind input. (a) Platform pitch average; (b) Platform pitch std; (c) Platform pitch maximum.

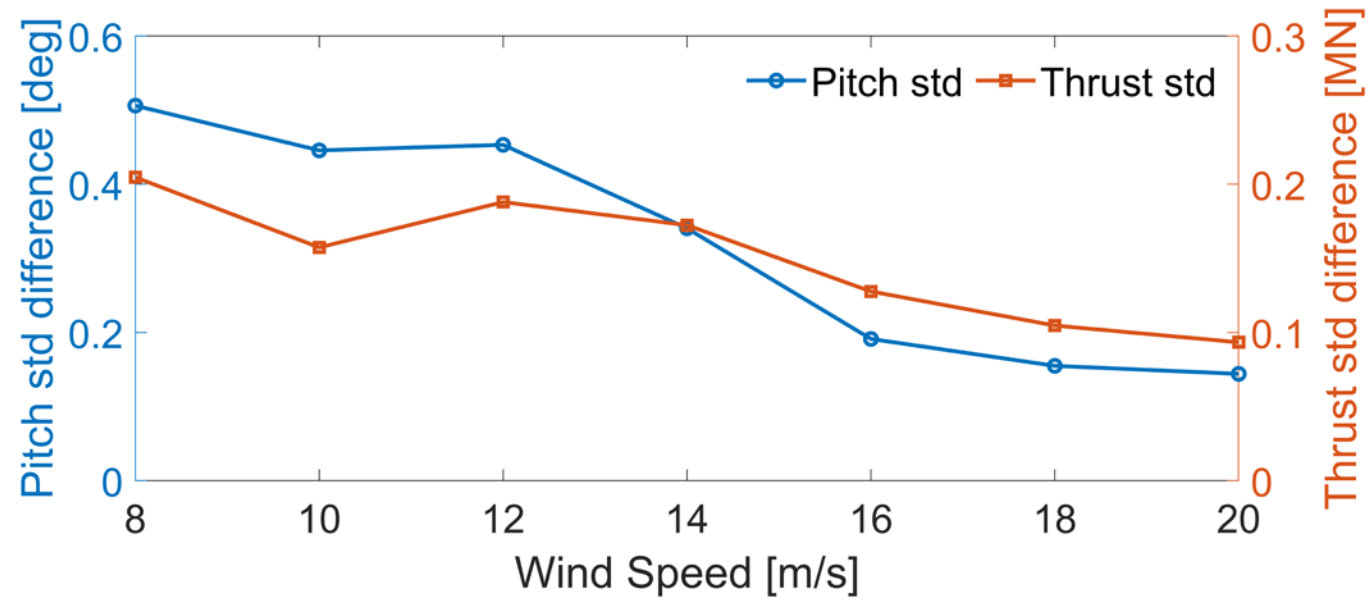

**Fig. 28.** The platform pitch and wind thrust std difference between turbulent and steady wind results with wind speed.

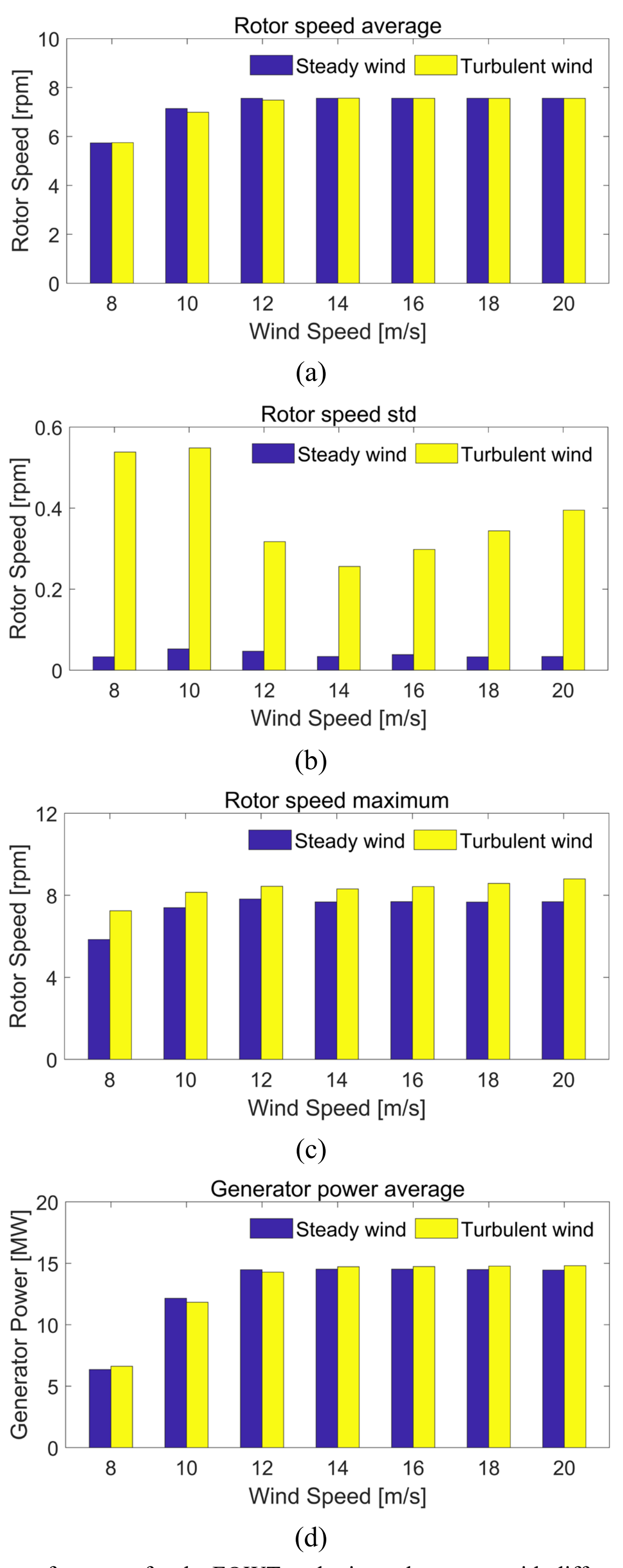

**Fig. 29.** Rotor performance for the FOWT under irregular waves with different wind input.
(a) Rotor speed average; (b) Rotor speed std; (c) Rotor speed maximum; (d) Generator power average.

#### *4.2.4. Comparison with conventional controllers*

In this section, we compare the proposed modified control strategy with several conventional control strategies, which determine the steady-state operating point based on the inflow wind speed instead of relative wind speed. The first one is the conventional baseline control strategy with $\omega_{\text{des}}$ = 0.3 rad/s larger than pitch natural frequency of the FOWT (around 0.22 rad/s). According to Abbas et al. (2022), a detuning control strategy with $\omega_{\mathrm{des}}$ = 0.2 rad/s is recommended as the second control strategy, while a smaller $\omega_{\mathrm{des}}$ of 0.04 rad/s is used as the third one.

Table 5 shows the results under steady wind with $U_{wind}$ = 12 m/s and regular waves with H = 5.0 m and T = 10 s. The conventional baseline control with a relatively large gain yields the largest platform pitch and rotor speed fluctuations due to the negative damping effect. The detuning strategy mitigates this effect, reducing the platform pitch and rotor speed fluctuations. However, the present modified control strategy has the best performance in reducing both platform pitch and rotor speed fluctuations for this case. Similar trends can also be seen in Table 6 for a turbulent wind and irregular wave test ($\bar{u}_{\mathrm{hub}}$ = 12 m/s, $H_s$ = 8.5 m and $T_p$ = 13.1 s).

**Table 5.** Comparison of results between different control strategies under steady wind ($U_{wind}$ = 12 m/s) and regular waves (H = 5.0 m and T = 10 s).

| Control strategy | Platform pitch (deg) | | Rotor speed (rpm) | |
|---|---|---|---|---|
| | max | std | max | std |
| Conventional baseline | 9.8145 | 3.2797 | 8.5914 | 0.6290 |
| Detuning ($\omega_{\mathrm{des}}$ = 0.2) | 5.7532 | 0.5251 | 7.7520 | 0.0912 |
| Detuning ($\omega_{\mathrm{des}}$ = 0.04) | 5.2048 | 0.4560 | 7.6975 | 0.0972 |
| Present modified | 5.0561 | 0.4771 | 7.5995 | 0.0241 |

**Table 6.** Comparison of results between different control strategies under turbulent wind ($\bar{u}_{\mathrm{hub}}$ = 12 m/s) and irregular waves (Hs = 8.5 m and $T_p$ = 13.1 s).

| Control strategy | Platform pitch (deg) | | Rotor speed (rpm) | |
|---|---|---|---|---|
| | max | std | max | std |
| Conventional baseline | 11.1719 | 3.0891 | 9.2586 | 0.6895 |
| Detuning ($\omega_{\mathrm{des}}$ = 0.2) | 9.0971 | 2.1834 | 8.6931 | 0.4923 |
| Detuning ($\omega_{\mathrm{des}}$ = 0.04) | 7.8546 | 1.2091 | 8.7469 | 0.3749 |
| Present modified | 7.2477 | 0.9530 | 8.4408 | 0.3172 |

## 5. Conclusions

A coupled aero-hydro-mooring-servo model with high computational efficiency is developed in this paper for the dynamic analysis of FOWTs in realistic operating environment. The coupled model combines the aerodynamic model OREGEN_BEMT, the hydrodynamic model OREGEN_X, the quasi-static mooring model or dynamic lumped-mass model MoorDyn, and multiple options of rotor control algorithms. A modified control strategy is proposed with a novel gain-scheduling technique for platform motion feedback to mitigate the negative damping effect on platform motions. In addition, the rotor dynamics is decoupled from the platform dynamics to obtain better rotor operating performance. For the IEA 15 MW bottom-fixed turbine, the steady-state operating points are presented under steady wind and verified by comparing with the NREL report (Gaertner et al., 2020). In turbulent wind simulations, we find that using the rotor-disk-averaged wind speed for control is beneficial to reducing fluctuations of wind thrust, generator power, rotor speed and blade pitch angle compared with using the hub-height wind speed. For the 15 MW FOWT, the negative damping phenomenon when using the baseline control is demonstrated through a range of steady uniform and turbulent wind conditions and wave scenarios including still water, regular and irregular waves. The effects of different control strategies on the platform response and rotor performance are extensively investigated. It indicates that the modified control strategy eliminates the negative damping effect on the platform pitch at above-rated wind speeds while maintaining small variations in rotor speed, showing the generally best performance among the three control strategies. Though the constant $k_\beta$ control can also eliminate the negative damping effect, it produces much larger fluctuations in rotor speed and causes larger overspeed exceeding the safety threshold of 20% at large wind speeds. Compared to steady wind, turbulent wind yields significantly larger low-frequency platform responses and increases the maximum rotor speed by 7.9% to 23.7% for the wind speeds considered here with the modified control strategy.

This study enhances our understanding of the negative damping phenomenon for FOWTs with the baseline control, and provides insights into the effects of turbulent wind and different control strategies on the dynamic responses and rotor performance of FOWTs in realistic offshore environments. Note that the wind turbine blades are assumed rigid and the aeroelastic effect is not considered here. Future research will incorporate the effect of blade deformations by coupling with a structural model. The present study may be regarded as a benchmark without this complication. Advanced aerodynamic models, like the CFD-based actuator line method, may be coupled with the present model to analyse the wake characteristics of FOWTs with rotor control. More advanced control strategies such as the individual blade pitch control, platform control using actuators and optimal control approaches will be considered in the future to further improve the platform stability, structural loading and power regulation of FOWTs.

## Acknowledgements

The authors gratefully acknowledge the financial support of EPSRC project FOWT-Control (Grant No. EP/W009854/1).